\documentclass[useAMS,usenatbib]{mn2e}
\usepackage{myaasmacros}
\usepackage{graphicx}
\usepackage{ulem}
\usepackage{color}
\usepackage[table]{xcolor}

\def\ltsima{$\; \buildrel < \over \sim \;$}
\def\simlt{\lower.5ex\hbox{\ltsima}}   
\def\gtsima{$\; \buildrel > \over \sim \;$}
\def\simgt{\lower.5ex\hbox{\gtsima}}
\newcommand\bcite[1]{\citeauthor{#1} \citeyear{#1}}

\def\unabla{{\bf \nabla}}
\def\uv{{\bf v}}

\def\urij{{\bf r}_{ij}}
\def\ur{{\bf r}}

\def\mtij{m_j}
\def\muE{|{\bf E}_0|}
\def\uE{{\bf E}_0}
\def\uEi{{\bf E}_{0,i}}

\def\wijtild{\overline{W}_{ij}}

\def\wij{W_{ij}}

\def\Mmv{{\bf V}}

\def\dx{x_{ij}}
\def\dy{y_{ij}}
\def\dz{z_{ij}}

\def\dthr{\mathrm{d}^3{ r}}

\def\Ron{RHA10}
\def\SPHG{SPHS}

\newlength{\myfix}
\title[SPH with a higher order dissipation switch]{SPHS: Smoothed Particle Hydrodynamics with a higher order dissipation switch}

\author[Read \& Hayfield]{J. I. Read$^{1,2}$\thanks{E-mail: justin.inglis.read@gmail.com} \& T. Hayfield$^{1,3}$\\
  $^1$Institute for Astronomy, Department of Physics, ETH Z\"urich, Wolfgang-Pauli-Strasse 16, CH-8093 Z\"urich, Switzerland\\
  $^2$Department of Physics and Astronomy, University of Leicester, University Road, LE1 7RH Leicester, UK\\
  $^{3}$Max-Planck-Institut f\"ur Astronomie, K\"onigstuhl 17, D-69117 Heidelberg, Germany\\
}

\begin{document}

\maketitle

\begin{abstract}
We present a novel implementation of Smoothed Particle Hydrodynamics (\SPHG) that uses the spatial derivative of the velocity divergence as a higher order dissipation switch. Our switch -- which is second order accurate -- detects flow convergence {\it before} it occurs. If particle trajectories are going to cross, we switch on the usual SPH artificial viscosity, as well as conservative dissipation in all advected fluid quantities (for example, the entropy). The viscosity and dissipation terms (that are numerical errors) are designed to ensure that all fluid quantities remain single-valued as particles approach one another, to respect conservation laws, and to vanish on a given physical scale as the resolution is increased. \SPHG\ alleviates a number of known problems with `classic' SPH, successfully resolving mixing, and recovering numerical convergence with increasing resolution. An additional key advantage is that -- treating the particle mass similarly to the entropy -- we are able to use multimass particles, giving significantly improved control over the refinement strategy. We present a wide range of code tests including the Sod shock tube, Sedov-Taylor blast wave, Kelvin-Helmholtz Instability, the `blob test', and some convergence tests. Our method performs well on all tests, giving good agreement with analytic expectations.
\end{abstract}


\section{Introduction}\label{sec:introduction}
Smoothed Particle Hydrodynamics (SPH) is now widely used in almost all areas of theoretical astrophysics \citep{1977MNRAS.181..375G,1977AJ.....82.1013L,1992ARA&A..30..543M}. Its popularity has been largely driven by its Lagrangian nature that makes it manifestly Galilean invariant and geometry-free; its ease of implementation; and the fact that it couples naturally to tree-gravity solvers that are currently the most efficient method for solving gravity (\bcite{1992ARA&A..30..543M}; \bcite{2005astro.ph..7472P}; \bcite{2009NewAR..53...78R}; \bcite{2010ARA&A..48..391S}; \bcite{2000ApJ...536L..39D}; \bcite{2011EPJP..126...55D}).

There are many different flavours of SPH used in the literature reflecting the above broad range of applications. The most common -- that we shall call `classic' SPH -- is the fully conservative SPH implemented in the standard release of the {\tt GADGET-2} code \citep{2002MNRAS.333..649S,2005MNRAS.364.1105S}\footnote{Slightly different implementations of this algorithm are also used in the literature, for example in the {\tt Gasoline} code \bcite{2004NewA....9..137W}, and the {\tt Hydra} code \bcite{1995ApJ...452..797C}. These are sufficiently similar to also be called `classic' SPH.}. Although classic SPH remains a powerful numerical tool for solving the fluid equations, it suffers from slow numerical convergence (\bcite{2010ARA&A..48..391S}), and a spurious surface tension at phase boundaries that inhibits fluid mixing \citep[see e.g.][]{1996PASA...13...97M,1999Dilts,2001MNRAS.323..743R,2006astro.ph.10051A,2008MNRAS.387..427W,2007arXiv0709.2772P,2009arXiv0906.0774R,2009arXiv0901.4107S}.

In recent work, we demonstrated that mixing in classic SPH fails for two distinct reasons (\bcite{2009arXiv0906.0774R}; \Ron). The first is a leading order error in the momentum equation, previously identified by \citet{1996PhDMorris} and \citet{1999Dilts}, that we called $\muE$. This can grow by orders of magnitude at flow boundaries, delaying the onset of instabilities. The second is a pressure discontinuity at flow boundaries, previously identified by \citet{2001MNRAS.323..743R}, \citet{2007arXiv0709.2772P} and \citet{2008MNRAS.387..427W}, that we called the local mixing instability (LMI)\footnote{It is an instability since, even if we start in pressure equilibrium, an infinitesimal perturbation will cause a pressure discontinuity to form.}. This leads to a large force error which manifests as a spurious surface tension. Both problems must be solved in order for mixing between fluids of different density or entropy to proceed correctly (see also \S\ref{sec:errors} and \S\ref{sec:cross} in this paper).

In \Ron\ we presented some simple proof-of-concept solutions to both of these problems. We cured the LMI by using a weighted density estimate first proposed by \citet{2001MNRAS.323..743R}, and we showed that $|\uE|$ can be made arbitrarily small by brute-force so long as the method is stable to large neighbour number (this required introducing some new kernels). However, our resulting Optimised Smoothed Particle Hydrodynamics (OSPH) method required a neighbour number that scales linearly with the density contrast on the kernel scale. The OSPH pressure estimator is also biased in regions of the flow where entropy gradients are large. This leads to poor performance in strong blast wave tests (we demonstrate this in Appendix \ref{sec:pestimate}). 

The above problems with SPH have led to a welcome proliferation of new Lagrangian or pseudo-Lagrangian techniques in the literature, including a moving-mesh code \citep{2009arXiv0901.4107S}, flux-based particle methods \citep{2010arXiv1006.4159G}, SPH using a Riemann solver \citep{2002JCoPh.179..238I,2010MNRAS.403.1165C, 2011arXiv1105.1344M}, and SPH using a Voronoi tessellation for the densities \citep{2010MNRAS.406.2289H}. It has also led to an exploration of improved flavours of SPH that add additional dissipation terms to mitigate the surface tension effect, and use switches to reduce the dissipation away from flow boundaries and shocks\footnote{Actually, some of these SPH flavours have been in use in the literature for quite some time (see e.g. \bcite{1997JCoPh.136...41M}).} (e.g. \bcite{2007arXiv0709.2772P}; \bcite{2009arXiv0902.4002K}; \bcite{2010JCoPh.229.8591R}; \bcite{2010MNRAS.408..669C}; \bcite{2011arXiv1111.1255P}).

In this paper, we present a new flavour of SPH -- \SPHG\ -- that has the mixing performance of OSPH, but does not introduce prohibitive numerical cost. As in OSPH, we use a larger than normal neighbour number with a correspondingly higher order and stable kernel to reduce the force errors. However, instead of the expensive OSPH pressure estimator, we introduce a higher order dissipation switch to ensure that all fluid quantities are smooth by construction. We show that these simple changes to the SPH algorithm lead to converged results with increasing resolution, and excellent performance across a wide range of code tests. Our dissipation switch also allows us to use multimass SPH particles. SPHS is useful for any astrophysics application involving multiphase fluid flow (e.g. resolving the ISM in galaxy discs), or where the use of multimass particles would be advantageous.

This paper is organised as follows. In \S\ref{sec:sfh}, \S\ref{sec:errors} and \S\ref{sec:cross} we present the \SPHG\ method. In \S\ref{sec:timestep}, we discuss our timestep criteria and multi-stepping scheme. In \S\ref{sec:implement}, we describe our implementation of \SPHG\ in the {\tt GADGET-2} code \citep{2005MNRAS.364.1105S}. In \S\ref{sec:tests}, we present a suite of tests for our new method that demonstrate that it can successfully model shocks, boundary instabilities, and shear flows. We also check that it conserves momentum, energy and mass and discuss the numerical performance of the code. Finally, in \S\ref{sec:conclusions}, we present our conclusions. 

\section{The \SPHG\ equations of motion}\label{sec:sfh}

In this paper, we consider solving the Euler equations in the absence of sinks or sources in the Lagrangian `entropy form' \citep{2002MNRAS.333..649S}: 

\begin{equation}
\frac{d\rho}{dt} = -\rho\unabla\cdot\uv
\label{eqn:euler1}
\end{equation}
\begin{equation}
\frac{d\uv}{dt}  = -\frac{\unabla P}{\rho}
\label{eqn:euler2}
\end{equation}
\begin{equation}
A = \mathrm{const.}
\label{eqn:euler3ent}
\end{equation}
closed by the ideal gas equation of state: 

\begin{equation}
P = A(s) \rho^{\gamma}
\label{eqn:enteulerstate}
\end{equation}
where $\gamma, \rho, v$ and $A$ are the adiabatic index, density, velocity and specific `entropy function' of the flow, respectively. The function $A(s)$ is a monotonic function of the specific entropy $s$. For adiabatic flow in the absence of sinks or sources, $A$ is conserved. Thus equation \ref{eqn:euler3ent} implicitly solves the energy equation. If required, the specific internal energy can be calculated from $A$ and $\rho$ as:  

\begin{equation}
u  = \frac{A \rho^{\gamma - 1}}{\gamma - 1}
\end{equation}
Note that often $A$ is referred to as the `entropy' when really it is a monotonic function of the specific entropy. From here on we will adopt this convention also. 

We use the discrete form of the above equations as in \Ron: 

\begin{equation}
\rho_i = \sum_j^N m_j \wij(|{\bf r}_{ij}|,h_i)
\label{eqn:sphcont}
\end{equation}
\begin{equation}
\frac{d\uv_i}{dt} = -\sum_j^N \frac{m_j}{\rho_i\rho_j} \left[P_i + P_j\right] \nabla_i \overline{W}_{ij}
\label{eqn:sphmoment}
\end{equation}
\begin{equation}
P_i = A_i \rho_i^\gamma
\label{eqn:sphstate}
\end{equation}
where $m_i$ is the mass of particle $i$; ${\bf r}_{ij} = {\bf r}_j - {\bf r}_i$; $\overline{W}_{ij} = \frac{1}{2}\left[W_{ij}(h_i) + W_{ij}(h_j)\right]$; and $W$ is a is a symmetric kernel that obeys the normalisation condition:
\begin{equation}
\int_{V} W(|\ur-\ur'|,h) \dthr' = 1
\label{eqn:normw}
\end{equation}
and the property (for smoothing length $h$):
\begin{equation}
\lim_{h\rightarrow 0} W(|\ur-\ur'|,h) = \delta(|\ur-\ur'|)
\end{equation}
Note that we do not explicitly solve the continuity equation nor the energy equation. The continuity equation is implicitly solved by equation \ref{eqn:sphcont} since its time derivative satisfies a discrete form of equation \ref{eqn:euler1} \citep[see e.g.][]{2005astro.ph..7472P}. The energy equation is implicitly solved by advecting the entropy function $A_i = \mathrm{const.}$ along with the particles \citep{2002MNRAS.333..649S}. 

We use a variable smoothing length $h_i$ as in \cite{2002MNRAS.333..649S} that is adjusted to obey the following constraint equation: 

\begin{equation}
\frac{4\pi}{3} h_i^3 n_i = N_n \qquad ;\,\mathrm{with} \qquad n_i = \sum_j^N W_{ij}
\label{eqn:fixedmass}
\end{equation}
where $N_n$ is the typical neighbour number (the number of particles inside the smoothing kernel, $W$). The above constraint equation gives fixed mass inside the kernel if particle masses are all equal. 

The above equations of motion manifestly conserve momentum, mass and entropy. They do not manifestly conserve energy, but the energy conservation is still extremely good as we will show in \S\ref{sec:conservation}. A fully conservative form of SPH can be constructed by replacing equation \ref{eqn:sphmoment} with equation \ref{eqn:sphmomentcons} (see Appendix \ref{sec:fullcons} and \bcite{2002MNRAS.333..649S}). However, as shown in \Ron, this leads to a larger truncation error in the momentum equation. In Appendix \ref{sec:fullcons}, we show that -- for the test problems presented in this paper -- the fully conservative form gives only a modest improvement in energy conservation while introducing significantly more diffusion for multiphase test problems. For this reason, we use the above set of equations as our default choice for \SPHG. 

So far, the above equations are very similar to classic SPH and thus will suffer from both the $\muE$ error and the LMI problems described in \S\ref{sec:introduction}. We now address each of these problems in turn in sections \S\ref{sec:errors} and \S\ref{sec:cross}.

\section{Errors \& convergence}\label{sec:errors} 

The first problem with classic SPH is the $\muE$ error in the momentum equation (\Ron). While this is minimised by using equation \ref{eqn:sphmoment}, the error is still present in \SPHG. To see this, we can Taylor expand $P_j$ about $P_i$ in equation \ref{eqn:sphmoment} to obtain\footnote{Note that this assumes that the pressures are smooth and therefore differentiable. In classic SPH, this is not guaranteed. However, in \SPHG, we add dissipation terms to ensure that this is the case. We discuss these in detail in \S\ref{sec:cross}.}: 

\begin{equation}
\frac{d\uv_i}{dt} \simeq  -\frac{2P_i}{h_i \rho_i}\sum_j \frac{m_j}{\rho_j}\unabla_i^x \overline{W}_{ij} - \frac{\left(\Mmv_i \unabla_i\right) P_i}{\rho_i}+ O(h) 
\label{eqn:taylor}
\end{equation}
where ${\bf V_i}$ is a matrix that approximates the identity matrix, and $\unabla_i^x = h_i \unabla$ is a dimensionless gradient operator. The left term in equation \ref{eqn:taylor} defines the dimensionless $\uEi$ error: 

\begin{equation}
\uEi = 2\sum_j^N \frac{m_j}{\rho_j} \unabla_i^x \overline{W}_{ij}
\end{equation}
Taking the limit of infinite kernel sampling (and equating $m_j/\rho_j$ with a volume element $dV$), we see that $\uEi$ is a discrete approximation to the volume integral:

\begin{equation}
\uEi \simeq 2 \int_V dV \unabla^x W = 0
\end{equation}
which is zero because $\unabla^x W$ is antisymmetric. 

Although $\uEi$ should be approximately zero, it is problematic because it appears in equation \ref{eqn:taylor} at order $h_i^{-1}$. Formal convergence then requires that $\uEi$ shrinks faster than $h_i$. This can be tricky to ensure and depends intimately upon the choice of kernel $W$ employed. A popular choice is the cubic spline (CS) kernel: 

\begin{equation}
W = \frac{8}{\pi h^3}\left\{\begin{array}{lr}
 1 - 6x^2 + 6x^3 & 0 < x \le \frac{1}{2}\\
2(1-x)^3 & \frac{1}{2} < x \le 1\\ 
0 & \mathrm{otherwise} \end{array}\right.
\label{eqn:cubicspline}
\end{equation}
where $x=r/h$ and, as written above, $h$ defines the kernel edge {\it not} its resolving power. (The two are not the same as can be readily understood by considering a Gaussian kernel. This has an infinite edge, but a resolving power given by $\sim$ its scale length.) 

\begin{center}
\begin{figure}
\begin{center}
\includegraphics[height=0.49\textwidth]{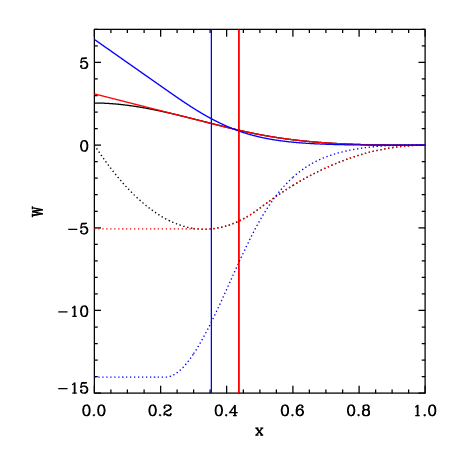}
\caption{The CS (black), CT (red) and HOCT4 (blue) kernels and their first derivatives (dotted lines). The vertical lines mark the half mass radii for each kernel. For the CS and CT kernels the half mass radii overlap on this plot.}
\label{fig:kernel_plot}
\end{center}
\end{figure}
\end{center}

Now, it is tempting to increase the kernel sampling simply by stretching $h$ for the CS kernel. However, this is a bad idea for two reasons. Firstly, it introduces bias into the density estimate, spoiling convergence. Secondly, the CS kernel is not stable to large neighbour number. As $h$ is increased, the particles clump on the kernel scale and the sampling is not significantly improved. For these reasons, in \Ron\ we proposed a new class of kernels that can be used to achieve convergence. The lowest order of these was the CT kernel: 

\begin{equation}
W = \frac{N}{h^3}\left\{\begin{array}{lr}
\left(-12\alpha + 18\alpha^2\right)x + \beta & 0 < x \le \alpha \\
 1 - 6x^2 + 6x^3 & \alpha < x \le \frac{1}{2}\\
2(1-x)^3 & \frac{1}{2} < x \le 1\\ 
0 & \mathrm{otherwise} \end{array}\right.
\label{eqn:ctkern}
\end{equation}
where $\beta = 1 + 6\alpha^2 - 12\alpha^3$, $N
= 8/[\pi \left(6.4\alpha^5 - 16\alpha^6 + 
  1\right)]$; and $\alpha = 1/3$. This has spatial resolution similar to the CS kernel but is stable to larger neighbour numbers. The next highest order was the HOCT4 kernel: 

\begin{equation}
W = \frac{N}{h^3}\left\{\begin{array}{lr}
Px + Q & 0 < x \le \kappa \\
(1-x)^4 + (\alpha - x)^4 + (\beta-x)^4 & \kappa < x \le \beta \\
(1-x)^4 + (\alpha - x)^4 & \beta < x \le \alpha\\
(1-x)^4 & x \le 1 \\
0 & \mathrm{otherwise} \end{array}\right.
\label{eqn:hoctkern}
\end{equation}
with $N = 6.515$, $P=-2.15$, $Q=0.981$, $\alpha = 0.75$, $\beta = 0.5$ and $\kappa = 0.214$. The CS, CT and HOCT4 kernels and their first derivatives are shown in Figure \ref{fig:kernel_plot}.

We demonstrated in \Ron\ that the HOCT4 kernel is stable for 442 neighbours on a lattice, while having a similar spatial resolution to the CT or CS kernel with 128 neighbours. This can be partially understood just from the half mass radii of these two kernels (Figure \ref{fig:kernel_plot}). If we assume that the resolution scale {\it is} the half mass radius, then 442 neighbours for the HOCT4 kernel is equivalent to $\sim 240$ neighbours for the CS kernel. However, in \Ron, we asses the spatial resolution more carefully by comparing instead the ability for these two kernels to correctly resolve the sound speed of linear waves. This is what leads to the conclusion that the HOCT4 kernel with 442 neighbours has similar resolving power to the CS with 128. Compared to `classic' SPH with 42 neighbours, the HOCT4 kernel with 442 neighbours has, therefore, a poorer spatial resolution of only a factor $\sim (128/42)^{1/3} \sim 1.5$. We will use the HOCT4 kernel with 442 neighbours for our default SPHS scheme. The CT kernel with 128 neighbours will be used only for convergence testing. 

As pointed out by \citet{2010arXiv1012.1885P}, the CT and HOCT4 kernels have slightly larger density error than the more standard CS kernel. However, the effect is small (see Table \ref{tab:denerror}). For glass particle distributions with 128 neighbours, the CT kernel gives a density error $\sim 2\%$ larger than the CS kernel, while the HOCT4 kernel with 442 neighbours is only $\sim 1$\% worse\footnote{It is not clear, given these results, why \citet{2010arXiv1012.1885P} argue that the density error is prohibitive for the CT kernel. Most likely it is because the error {\it is} very large when only 32 neighbours are used. For larger neighbour numbers, however, the performance of the CT and HOCT4 kernels is acceptable.}. (Note that the error can be very large if too few neighbours are used: higher order kernels require more neighbours to be adequately sampled.)

\begin{table}
\center
\caption{Density errors for a selection of SPH kernels applied to a constant density box. The columns give: kernel type; neighbour number; lattice configuration (glass or simple cubic); and the median/5\%/95\% recovered density to two significant figures (the true density is $\rho_\mathrm{true} = 1.00$).}
\begin{tabular}{cccccc}
\hline
Kernel & $N_n$ & Lattice & $\rho$ (5\%) & $\rho$ (median) & $\rho$ (95\%)\\
\hline 
\cellcolor[gray]{1.0} CS & \cellcolor[gray]{1.0} 32 & \cellcolor[gray]{1.0} simple & \cellcolor[gray]{1.0} 1.00 & \cellcolor[gray]{1.0} 1.00 & \cellcolor[gray]{1.0} 1.00\\
\cellcolor[gray]{0.9} CS & \cellcolor[gray]{0.9} 128 & \cellcolor[gray]{0.9} simple & \cellcolor[gray]{0.9} 1.00 & \cellcolor[gray]{0.9} 1.00 & \cellcolor[gray]{0.9} 1.00\\
\cellcolor[gray]{1.0} CS & \cellcolor[gray]{1.0} 32 & \cellcolor[gray]{1.0} glass & \cellcolor[gray]{1.0} 1.02 & \cellcolor[gray]{1.0} 1.01 & \cellcolor[gray]{1.0} 1.01\\
\cellcolor[gray]{0.9} CS & \cellcolor[gray]{0.9} 128 & \cellcolor[gray]{0.9} glass & \cellcolor[gray]{0.9} 1.01 & \cellcolor[gray]{0.9} 1.00 & \cellcolor[gray]{0.9} 0.996\\
\cellcolor[gray]{1.0} CT & \cellcolor[gray]{1.0} 32 & \cellcolor[gray]{1.0} simple & \cellcolor[gray]{1.0} 1.07 & \cellcolor[gray]{1.0} 1.07 & \cellcolor[gray]{1.0} 1.07\\
\cellcolor[gray]{0.9} CT & \cellcolor[gray]{0.9} 128 & \cellcolor[gray]{0.9} simple & \cellcolor[gray]{0.9} 1.02 & \cellcolor[gray]{0.9} 1.02 & \cellcolor[gray]{0.9} 1.02\\
\cellcolor[gray]{1.0} CT & \cellcolor[gray]{1.0} 32 & \cellcolor[gray]{1.0} glass & \cellcolor[gray]{1.0} 1.08 & \cellcolor[gray]{1.0} 1.07 & \cellcolor[gray]{1.0} 1.06\\
\cellcolor[gray]{0.9} CT & \cellcolor[gray]{0.9} 128 & \cellcolor[gray]{0.9} glass & \cellcolor[gray]{0.9} 1.02 & \cellcolor[gray]{0.9} 1.02 & \cellcolor[gray]{0.9} 1.01\\
\cellcolor[gray]{1.0} HOCT4 & \cellcolor[gray]{1.0} 442 & \cellcolor[gray]{1.0} simple & \cellcolor[gray]{1.0} 1.01 & \cellcolor[gray]{1.0} 1.01 & \cellcolor[gray]{1.0} 1.01\\
\cellcolor[gray]{0.9} HOCT4 & \cellcolor[gray]{0.9} 442 & \cellcolor[gray]{0.9} glass & \cellcolor[gray]{0.9} 1.02 & \cellcolor[gray]{0.9} 1.01 & \cellcolor[gray]{0.9} 1.00\\
\hline
\label{tab:denerror}
\end{tabular}
\end{table}

In summary, formal convergence in SPH is somewhat subtle; it requires several important criteria to be satisfied: 

\begin{enumerate} 

\item increased particle number, $N$;

\item increased neighbour number, $N_n$ to ensure that $\uEi$ shrinks faster than $h_i$;

\item a higher order kernel to maintain spatial resolution; and 

\item a kernel that is stable to clumping/banding for the above choice of $N_n$. 

\end{enumerate} 
The CT kernel with 128 neighbours and the HOCT4 kernel with 442 neighbours thus provide a convergent kernel pair that satisfy the above criteria. We will demonstrate this in \S\ref{sec:vortex}. (Note that the above is simply what is required {\it formally}. It may be that for a given numerical problem $\uEi$ shrinks faster than $h_i$ without any need to raise the neighbour number. This cannot be guaranteed in general, however.)

The above convergence criteria -- constructed simply to ensure that $\uEi$ shrinks faster than $h_i$ -- seem rather laborious. Given that we know a priori what $\uEi$ is for each particle, it is tempting to simply factor it out of the momentum equation as follows:  

\begin{equation}
\frac{d\uv_i}{dt} = -\sum_j^N \frac{\mtij}{\rho_i \rho_j}\left[P_j - P_i \right] \unabla_i \overline{W}_{ij} - \frac{P_i}{h_i \rho_i}\uEi
\label{eqn:limit}
\end{equation}
The left term is now a higher order momentum equation\footnote{The left term in equation \ref{eqn:limit} is remarkably similar to the momentum equation discussed and proposed recently by \citet{2011MNRAS.tmp..196A}. Such an equation has been proposed several times in the literature before (e.g. \bcite{1995astro.ph..3124M}). As was pointed out in \Ron, it improves mixing in SPH because it manifestly removes the $\uEi$ error. However, if we only subtract $\uEi$ and do nothing else, the method will fail because of a lack of momentum conservation in strong shocks, and because nothing has been done to mitigate the LMI.}. It gives zero force for constant pressure by construction, unlike equation \ref{eqn:sphmoment}. And, it should give much simpler convergence -- no longer requiring increased neighbour number, or careful kernel choice (indeed we will demonstrate this in \S\ref{sec:vortex}). However, these advantages come at a price. As pointed out in \Ron, notice that the left term in equation \ref{eqn:limit} is {\it symmetric} in $i$ and $j$ inside the sum. This means that momentum is no longer conserved between particle pairs. This lack of manifest momentum conservation becomes a problem in strong shocks \citep{1996PASA...13...97M}. We will discuss subtracted-$\uEi$ momentum equations and their potential for creating higher order SPH-like methods in a separate paper (Hayfield \& Read 2011, in prep.).

\section{Convergent flow \& dissipation}\label{sec:cross} 

\begin{center}
\begin{figure}
\begin{center}
\includegraphics[height=0.3\textwidth]{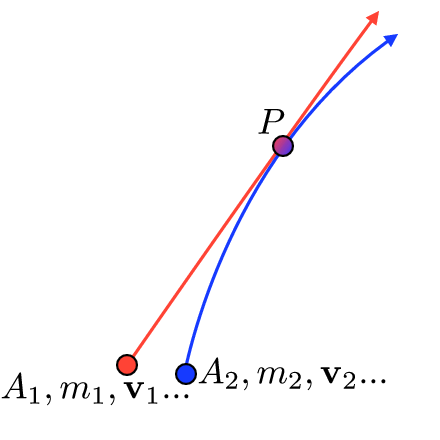}
\vspace{-5mm}
\caption{A schematic representation of two SPH particles approaching one another in a convergent flow. The two particles carry discretely different advected quantities with them: entropy, $A_{1,2}$, mass $m_{1,2}$, velocity ${\bf v}_{1,2}$ etc. Apart from their densities that are manifestly smooth (c.f. equation \ref{eqn:sphcont}), all other fluid quantities become multi-valued at point $P$. Thus, we must detect when this situation is going to occur and add dissipation terms in {\it all} fluid quantities to ensure that they remain single-valued throughout the flow.}
\label{fig:trajectory_cross}
\end{center}
\end{figure}
\end{center}

The second problem with SPH is dealing with flow convergence -- a problem common to any Lagrangian scheme. SPH can be thought of as both a Monte-Carlo method and a method of characteristics. It is a Monte-Carlo method because a finite number of discrete particles are used to initially sample the fluid. However, from this moment onwards it a method of characteristics: the particles move along streamlines in the flow. The problem is that the particles represent large unresolved patches of the fluid. Unlike real infinitesimal points in a fluid flow, SPH particles can approach one another, as shown in Figure \ref{fig:trajectory_cross}. This leads to multivalued fluid quantities at the crossing point: multivalued momentum, entropy, mass, and any other fluid quantity that is advected with the particles. The only quantity that is not multivalued is the density since this is calculated by smoothing over a particles' nearest neighbours (see equation \ref{eqn:sphcont}). 

The problem of particles approaching one another was realised very early on in the development of SPH, and led to the introduction of artificial viscosity. This acts to make the momentum between particles single valued as they approach one another, while maintaining energy and momentum conservation. However, less appreciated in the literature is the need for similar dissipation terms in all other advected fluid quantities. This was recently highlighted by \citet{2007arXiv0709.2772P}. For example, if two particles approach one another with very different entropy, their pressures will become multivalued. This leads to a spurious repulsive force between the particles which inhibits mixing. In \Ron, we referred to this as the Local Mixing Instability (LMI). However, it can be thought of as a more general problem of multivalued fluid quantities arising when particles approach one another. 

There are two possible solutions to deal with multivalued pressures in SPH. In \Ron, we used an idea from \citet{2001MNRAS.323..743R} to manifestly smooth the pressures by using a modified `RT' density estimator: 

\begin{equation}
\rho_i = \sum_j^N \left(\frac{A_j}{A_i}\right)^{\frac{1}{\gamma}} m_j \wijtild
\label{eqn:sphcontrtent}
\end{equation}
which gives (via equation \ref{eqn:sphstate}): 

\begin{equation}
P_i = \left[\sum_j^N A_j^{\frac{1}{\gamma}} m_j \wijtild\right]^\gamma
\label{eqn:rtp}
\end{equation}
The above density estimator, combined with equations \ref{eqn:sphcont} and \ref{eqn:sphmoment}, defines the OSPH method derived in \Ron. 

The `RT' density estimator has the nice feature that it avoids multivalued pressures {\it by construction}. However, there is an associated cost. Consider the situation of large entropy contrasts on the kernel scale. Particles with $A_i \gg A_j$ will contribute essentially zero weight, reducing the effective kernel sampling. To maintain a constant $\muE$ error, we must then scale the neighbour number proportional to the entropy contrast on the kernel scale. This becomes prohibitively expensive for astrophysically important applications like strong blast waves. Here, OSPH gives significantly poorer performance than SPH for the same numerical cost. We demonstrate this in Appendix \ref{sec:pestimate}. 

For the above reasons, in this paper we take an approach more similar to \citet{2007arXiv0709.2772P}, but with a key difference. \citet{2007arXiv0709.2772P} presented dissipation switches designed to detect (and correct) multivalued pressures. However, once pressures are multivalued it is already too late. As demonstrated recently by \citet{2010MNRAS.408...71V}, once pressure blips form at flow boundaries, they cause pressure waves to propagate throughout the fluid. These damp the growth of surface instabilities and cause errors to propagate throughout the flow. To avoid this problem, we must detect when particles will approach one another {\it in advance}. We can then act to ensure that all fluid quantities (not just the pressure) will be single valued by the time the particles reach one another. This is the strategy we adopt here. 

To detect when particles will cross, we require an accurate flow convergence detector. We take an approach similar to \citet{2010MNRAS.408..669C}. \citet{2010MNRAS.408..669C} came up with the novel idea of using the time derivative of the velocity divergence to detect flow convergence in advance. They then switch on artificial viscosity to prevent particle inter-penetration. We use a similar idea, but consider instead the {\it spatial derivative} of the velocity divergence. As we will show, this has the advantage that we obtain an excellent estimate of the flow divergence and curl for free. 

\citet{2010MNRAS.408..669C} focus only on the artificial viscosity. Here, we use the same switch not just for the artificial viscosity, but for {\it all} artificial dissipation terms. (Recall that we require one of these for each advected fluid quantity.) 

We have some freedom in how to construct the flow convergence detector and the dissipation terms. However, both must satisfy a number of constraints in order for the scheme to produce converged results with increasing resolution: 

\begin{enumerate}

\item the switch must detect flow convergence {\it before it occurs}; 

\item the switch must be sufficiently robust (i.e. high order) as to not trigger randomly due to particle noise; 

\item the dissipation terms must respect conservation laws;

\item the dissipation terms must shrink on a given physical scale with increasing resolution; and 

\item the dissipation terms must not generate spurious pressure waves that propagate through the fluid. 

\end{enumerate}
These criteria guide our choices for the switch and the artificial dissipation terms that we describe in the following two subsections. The last point, in particular, is important. It is no good if our dissipation terms introduce more problems than they solve. They should act to make fluid quantities single valued wherever particles approach one another. But they should do this in a manner that respects conservation laws, is convergent, and does not lead to problems elsewhere in the flow. 

\subsection{A higher order convergence detector} 

We first describe our higher order flow convergence detector. Local flow convergence occurs wherever the velocity divergence is negative. This suggests that we should switch on dissipation terms if $\unabla \cdot {\bf v}_i < 0$ for a given particle. However, if we set the magnitude of the dissipation also using $\unabla \cdot {\bf v}_i$, then the dissipation will only switch on once the flow is converging, not before. To detect flow convergence in advance, we use instead the {\it spatial derivative} of $\unabla \cdot {\bf v}_i$ for the magnitude of our dissipation parameter $\alpha_{\mathrm{loc},i}$. This leads to the following dimensionless dissipation switch: 

\begin{equation}
\alpha_{\mathrm{loc},i} = \left\{
\begin{array}{lr}
\frac{h_i^2 |\unabla(\unabla \cdot {\bf v}_i)|}{h_i^2 |\unabla(\unabla \cdot {\bf v}_i)| + h_i |\unabla \cdot {\bf v}_i|+ n_s c_s} \alpha_\mathrm{max} & \unabla \cdot {\bf v}_i < 0 \\
0 & \mathrm{otherwise} 
\end{array}\right.
\label{eqn:alphalocv}
\end{equation}
where $\alpha_{\mathrm{loc},i}$ describes the amount of dissipation for a given particle in the range $[0,\alpha_\mathrm{max} = 1]$; and $n_s = 0.05$ is a `noise' parameter that determines the magnitude of velocity fluctuations that trigger the switch. Equation \ref{eqn:alphalocv} turns on dissipation if $\unabla \cdot {\bf v}_i < 0$ (convergent flow) and if the magnitude of the spatial derivative of $\unabla \cdot {\bf v}_i$ is large as compared to the local divergence (i.e. if the flow is going to converge). 

\begin{center}
\begin{figure*}
\hspace{3mm}\includegraphics[width=\textwidth]{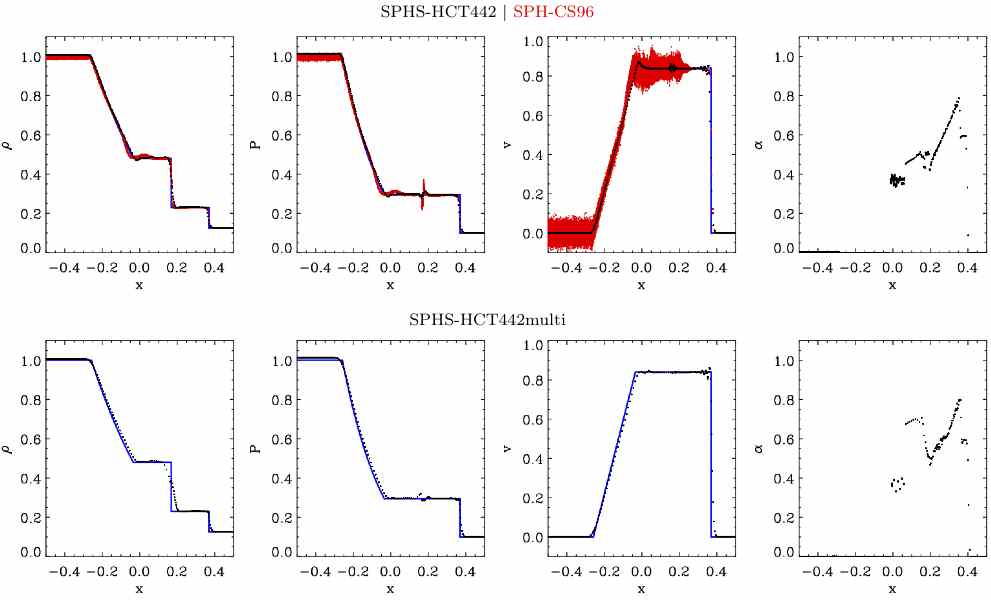}
\caption{3D Sod shock tube test results at $t=0.2$ in \SPHG\ (black) and SPH-CS96 (red). From left to right, the panels show: density; pressure; the magnitude of $v_x$ (the $x$-velocity component along the shock); and the dissipation switch $\alpha$ (only relevant for the \SPHG\ simulations). The blue line marks the analytic solution. Notice that the pressure blip at $x \sim 0.2$ is almost fully removed in \SPHG. The top panels show results for a single particle-mass simulation; the bottom for the same run with multimass particles in \SPHG\ on a uniform particle grid.}
\label{fig:sod}
\end{figure*}
\end{center}

In principle, the maximum dissipation parameter $\alpha_\mathrm{max}$ can be different for each fluid quantity. Our default in this paper is to use $\alpha_\mathrm{max} = 1$ for all fluid variables. We investigate the sensitivity of \SPHG\ to $\alpha_\mathrm{max}$ in Appendix \ref{sec:alphasen}.

As in \citet{2010MNRAS.408..669C}, we set the local dissipation to the above value instantaneously if $\alpha_i < \alpha_{\mathrm{loc},i}$:

\begin{equation}
\begin{array}{lr}
\alpha_i = \alpha_{\mathrm{loc},i} \hspace{4cm}& \alpha_i < \alpha_{\mathrm{loc},i}
\end{array}
\label{eqn:alpha}
\end{equation}
otherwise, $\alpha_i$ smoothly decays back to zero: 
\begin{equation}
\begin{array}{lr}
\dot{\alpha}_i = (\alpha_{\mathrm{loc},i} - \alpha_i)/\tau_i \hspace{1.5cm}& \alpha_\mathrm{min} < \alpha_{\mathrm{loc},i} < \alpha_i \\
\dot{\alpha}_i  = (\alpha_\mathrm{min} - \alpha_i)/\tau_i & \alpha_\mathrm{min} > \alpha_{\mathrm{loc},i}\\
\end{array}
\label{eqn:alphadot}
\end{equation}
where $\tau_i = h_i/v_{\mathrm{max},i}$ is the timescale for the decay; and $v_{\mathrm{max},i}$ is the maximum signal velocity \citep{2005MNRAS.364.1105S}:
\begin{equation}
v_{\mathrm{max},i} = \max_j\left[v_\mathrm{sig,{ij}}\right] 
\label{eqn:maxvsig}
\end{equation}
with
\begin{equation}
v_\mathrm{sig,{ij}} = c_i + c_j - 3 w_{ij}
\label{eqn:vsig}
\end{equation}
where $w_{ij} = \frac{{\bf v}_{ij} \cdot {\bf r}_{ij}}{|{\bf r}_{ij}|}$, and $c_i$ is the local sound speed at particle $i$.

The parameter $\alpha_\mathrm{min} = 0.2$ ensures that the dissipation parameter decays all the way back to zero once particles are no longer converging. 

\subsection{A higher order gradient estimator}\label{sec:grad} 

Our dissipation switch (equation \ref{eqn:alphalocv}) requires a good estimate of both the first and second derivatives of the velocity field. A noisy estimator will cause the limiter to trigger unnecessarily, leading to an overly diffusive method\footnote{Indeed, \citet{2010JCoPh.229.8591R} recently advanced the idea of using higher order gradients for their dissipation switch. They used a first order accurate gradient of the pressure, whereas we use the gradient of the velocity divergence (which is a second derivative of the velocity field).}. To achieve good quality gradients, we fit a second order polynomial to each of the fluid variables as in \citet{2003ApJ...595..564M}. The first and second derivatives then follow from the coefficients of the polynomial fit. The full 3D algorithm is given in Appendix \ref{sec:full}. Here we present a 1D version to illustrate the idea.  

We assume that a fluid variable, $q_i$, can be locally represented by a smooth second order polynomial: 

\begin{equation}
q_i = a_{0,i} + a_{1,i} x_{ij} + a_{2,i} x_{ij}^2 + O(h^3)
\end{equation}
where $x_{ij} = r_{ij} / h_i$. 

To determine the coefficients $a_{n,i}$, we then consider the matrix equation ${\bf M} {\bf a} = {\bf q}$: 

\begin{eqnarray}
\left[
\sum_j^N m_j W_{ij}
\left(\begin{array}{rrr}
1 & x_{ij} & x_{ij}^2 \\
x_{ij} & x_{ij}^2 & x_{ij}^3 \\
x_{ij}^2  & x_{ij}^3 & x_{ij}^4 \\
\end{array} \right)\right]
\left(\begin{array}{c}
a_{0,i}\\
a_{1,i}\\
a_{2,i}\\
\end{array}
\right)
 & = & \nonumber \\
  \sum_j^N m_j W_{ij}
\left(\begin{array}{c}
q_j\\
q_j x_{ij}\\
q_j x_{ij}^2\\
\end{array}
\right) & & \\
\end{eqnarray}
The matrix ${\bf M}$ and the vector ${\bf q}$ contain weighted moments that can be calculated in the usual way by summing over each particle's nearest neighbours. The vector ${\bf a}$ is then calculated by solving for the inverse of ${\bf M}$. The particle gradients at the position of the particle ($x_{ij} = 0$) then follow from $q_i'(0) = a_{1,i}$ and $q_i''(0) = 2 a_{2,i}$.

The above straightforwardly generalises to 3D and to vector fluid variables. For scalar variables in 3D we must solve a $10\times 10$ matrix inverse to obtain a 10 coefficient fit (see Appendix \ref{sec:full}):

\begin{eqnarray}
q_{ij}  & = & a_{0,i} + a_{1,i} x_{ij} + a_{2,i} y_{ij} + a_{3,i} z_{ij} + a_{4,i} x_{ij}^2 + a_{5,i} y_{ij}^2 + \nonumber \\
& &  a_{6,i} z_{ij}^2 + a_{7,i} x_{ij} y_{ij} + a_{8,i} x_{ij}z_{ij} + a_{9,i} y_{ij} z_{ij}  + \nonumber \\
& & O(h^3)
\label{eqn:polyfit}
\end{eqnarray}
where ${\bf x_{ij}} = \urij / h_i = [x_{ij}, y_{ij}, z_{ij}]$. 

Note that \citet{2003ApJ...595..564M} use these higher order gradients to actually move the fluid. This makes the method non-conservative, leading to problems in strong shocks. In \SPHG, we use these gradients instead to conservatively maintain fluid smoothness. 

Our dissipation switch manifestly satisfies our criteria (i) and (ii) outlined above. It detects flow convergence in advance, and it is accurate since it is based on a second order accurate expansion of the velocity field.  

Note that a second order polynomial is the lowest order that we could fit in order to obtain a second derivative. In principle, we could fit a third or fourth order polynomial thus further increasing the accuracy of the switch. However, this comes at quite significant cost. At third order, the size of the moment matrix increases from $10\times 10$ to $20\times 20$ and requires an additional 40 sums over the particles to be calculated and stored. Secondly, for the higher order moments to actually help, the neighbour number should be increased. Otherwise noise in the third moments could make the higher order gradient estimator poorer than the second order estimate. For these reasons, we stick to the second order scheme in this paper.

\subsection{The dissipation terms}\label{sec:dissipation}

\subsubsection{Artificial viscosity} 

We start with the familiar artificial viscosity. Here, we use the form as in \citet{Monaghan1997298} and \citet{2005MNRAS.364.1105S}: 

\begin{equation}
\dot{{\bf v}}_\mathrm{diss,i} = -\sum_j^N m_j \Pi_{ij} \unabla_i \wijtild
\label{eqn:viscmom}
\end{equation}
where: 
\begin{equation}
\Pi_{ij} = \left\{\begin{array}{cl}
-\frac{\overline{\alpha}_{ij}}{2} \frac{v_\mathrm{sig,{ij}} w_{ij}}{\overline{\rho}_{ij}} & \mathrm{if\,} {\bf v}_{ij} \cdot {\bf r}_{ij} < 0 \\
0 & \mathrm{otherwise} \\
\end{array}\right.
\label{eqn:pivisc}
\end{equation}
where $\overline{\alpha}_{ij} = \frac{1}{2}\left[\alpha_i + \alpha_j\right]$, and $v_\mathrm{sig,{ij}}$ and $w_{ij}$ are defined by equation \ref{eqn:vsig}. This must then generate entropy to ensure energy conservation: 
\begin{equation}
\dot{A}_\mathrm{diss,i} = -\frac{1}{2}\frac{\gamma-1}{\rho_i^{\gamma-1}} \sum_j^N m_j \overline{\alpha}_{ij} \Pi_{ij} {\bf v}_{ij} \cdot \unabla_i \wijtild
\label{eqn:visc}
\end{equation}
In addition, we use a Balsara-like switch to limit viscosity in shear flows \citep{balsaraphd89}. As in \citet{2010MNRAS.408..669C}, we apply this to our viscosity parameter $\alpha_i$, rather than directly to equation \ref{eqn:visc}. This is mathematically identical, but means that $\alpha_i$ represents the true viscosity of the flow. Thus, we multiply $\alpha_i$ by a suppression function given by: 

\begin{equation}
f_{\mathrm{Balsara},i} = \frac{|\nabla \cdot {\bf v}|_i}{|\nabla \cdot {\bf v}|_i + |\nabla \wedge {\bf v}|_i + 0.0001 c_{s,i}/h_i}
\label{eqn:balsara}
\end{equation}
where $c_{s,i}$ is the sound speed for particle $i$. Equation \ref{eqn:balsara} is identical to the usual Balsara switch, except that we use the higher order gradients derived in Appendix \ref{sec:full} to derive the divergence and curl of the velocity field.

Equations \ref{eqn:viscmom} and \ref{eqn:visc} satisfy our dissipation criteria (iii)-(v) outlined above. They respect energy and momentum conservation by construction; they act only on the kernel scale (and thus the viscosity will reduce at a given physical scale as the resolution is increased); and they introduce a numerical error only locally. 

\subsubsection{Entropy dissipation} 

For our dissipation in the entropy function variable $A_i$, we choose a form that explicitly conserves energy, similar to that proposed in \citet{2007arXiv0709.2772P}: 

\begin{equation}
\dot{A}_{\mathrm{diss},i} = 
\sum_j^N \frac{m_j}{\overline{\rho}_{ij}} \overline{\alpha}_{ij} {v}_{\mathrm{sig},ij}^p L_{ij} \left[A_i - A_j \left(\frac{\rho_j}{\rho_i}\right)^{\gamma-1}\right] K_{ij}
\label{eqn:entdiss}
\end{equation}
where $\overline{\rho}_{ij} = [\rho_i + \rho_j]/2$ is the symmetrised density; $K_{ij} = \hat{r}_{ij} \cdot \nabla_i W_{ij}$ is a symmetric smoothing kernel; $L_{ij}$ is a pressure limiter (of which more in a moment); and ${v}_{\mathrm{sig},ij}^p$ is similar to the signal velocity, but defined to be positive definite: 

\begin{equation}
v_\mathrm{sig,{ij}}^p = \left\{\begin{array}{cl}
c_i + c_j - 3 w_{ij} & \mathrm{if}\, 3w_{ij} < (c_i+c_j) \\
0 & \mathrm{otherwise}
\end{array}\right .
\label{eqn:vsigp}
\end{equation}
This modified signal velocity is chosen to give more dissipation to approaching particle pairs than receding particle pairs. However, unlike the viscosity where the dissipation is fully suppressed for receding pairs (c.f. equation \ref{eqn:pivisc}), we find that receding pairs still require some small entropy dissipation. This is because, while neighbouring particles can have discretely different velocities without serious repercussion (so long as they are not approaching one another), discretely different entropies inside the kernel will drive spurious pressure waves that affect the numerical solution everywhere. 

In fact, the above explains why adding some small entropy dissipation is preferable to doing nothing at all. The right amount of entropy dissipation will ensure smooth pressures and keep errors local. But the key is getting the `right amount'. If we are not careful, our dissipation terms can actually drive pressure waves and do more harm than good. To avoid this, we introduce a pressure limiter: 

\begin{equation}
L_{ij} = \frac{|P_i - P_j|}{P_i + P_j}
\label{eqn:plim}
\end{equation}
Note that, unlike the dissipation prescription presented in \citet{2007arXiv0709.2772P}, equation \ref{eqn:entdiss} poses no problem for simulations involving gravity. In hydrostatic equilibrium the entropy dissipation will vanish since the flow is non-converging and $\overline{\alpha}_{ij} = 0$.  

Equation \ref{eqn:entdiss} satisfies our dissipation criteria (iii)-(v) outlined above. It respects energy conservation by construction; acts only on the kernel scale; and -- through the pressure limiter -- does not propagate errors non-locally. 

\subsubsection{Mass dissipation (for multimass applications)}\label{sec:diffm}

Multimass SPH particles are very useful since they allow interesting regions of the flow to be simulated at significantly higher resolution (e.g. \bcite{1988MNRAS.231..515M}; \bcite{1993MNRAS.264..691M}). However, classic SPH runs into difficulties once particle masses are allowed to vary (see e.g. \bcite{2003physics...3112O}). The problems occur because, like the entropy, the particle masses are advected along with the particles. When particles approach one another, the masses become multivalued, driving a pressure wave at the mass interface. The problem is less severe than for the entropies because the masses are smoothed over in the equation of state (c.f. equations \ref{eqn:sphcont} and \ref{eqn:sphstate}). Nonetheless, large density contrasts realised with multimass particles are problematic. 

Some approaches to multimass SPH have been proposed in the literature. \citet{2003physics...3112O} suggest adapting the density estimate to ensure smooth pressures by construction -- an approach very similar to the multiphase SPH proposed by \citet{2001MNRAS.323..743R}. \citet{2002MNRAS.330..129K} suggest increasing the neighbour number at course-fine boundaries. This will act to smooth any pressure blips at the interface and is therefore also a viable solution.

\begin{center}
\begin{figure}
\begin{center}
\hspace{-6mm}
\includegraphics[height=0.49\textwidth]{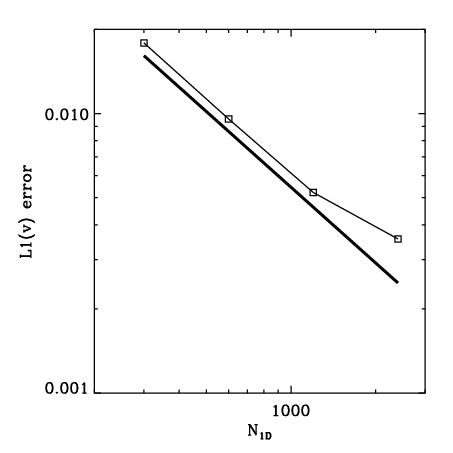}
\caption{Sod shock tube convergence test results. The x-axis gives the number of particles along the shock $N_{1D}$ (Figure \ref{fig:sod}, top, shows results for $N_{1D}=600$); the y-axis shows the binned velocity error in $x$-bins of width 0.01; and the thick black line shows a scaling of $N_{1D}^{-0.9}$ (the best-possible scaling is $N_{1D}^{-1}$).}
\label{fig:sod_converge}
\end{center}
\end{figure}
\end{center}

A key advantage of our approach is that we can treat {\it any} advected fluid quantity in the same manner as the entropy, above. This includes the particle masses, which allows us to consider a multimass SPH scheme that does not require raising the neighbour number at boundaries, or introducing a new density estimator. Treating the mass similarly to the entropy, above, we introduce a conservative pairwise mass dissipation: 

\begin{equation}
\dot{m}_{\mathrm{diss},i} = \sum_j^N \frac{\overline{m}_{ij}}{\overline{\rho}_{ij}} \overline{\alpha}_{ij} {v}_{\mathrm{sig},ij}^p  L_{ij} \left[m_i - m_j \right] K_{ij}
\label{eqn:mdiss}
\end{equation}
where $\overline{m}_{ij} = [m_i + m_j]/2$ is the symmetrised mass. Note that this symmetrised mass appears {\it only} in equation \ref{eqn:mdiss}, and not in the other dissipation terms. This difference follows from the fact that equation \ref{eqn:mdiss} must respect mass conservation, while equations \ref{eqn:entdiss} and \ref{eqn:visc} -- that describe the evolution of the {\it specific} entropy -- must respect energy conservation (see Appendix \ref{sec:massdisscons} for further details). 

As for the artificial viscosity, we must then add correction terms to ensure momentum and energy conservation. There is actually some freedom in how we choose to do this (see Appendix \ref{sec:massdisscons}). A simple approach is to ensure that each particle individually conserves its energy and momentum: 

\begin{equation}
\frac{d (m_i {\bf v}_i)}{dt} = \dot{m}_i{\bf v}_i  + m_i\dot{{\bf v}}_i = 0
\label{eqn:masscons}
\end{equation}
\begin{equation}
\frac{d E_i}{dt} = \dot{m}_i u_i + m_i \dot{u}_i + \frac{1}{2}\dot{m}_i{\bf v}_i\cdot {\bf v}_i + m_i {\bf v}_i \cdot\dot{{\bf v}}_i
\label{eqn:econs}
\end{equation}
where we recall that $u_i$ is the specific internal energy. Substituting equation \ref{eqn:masscons} into equation \ref{eqn:econs} then gives the correction terms for each particle: 

\begin{equation}
\dot{{\bf v}}_{\mathrm{diss},i} = - \frac{\dot{m}_{\mathrm{diss},i}}{m_i} {\bf v}_i
\label{eqn:vdotm}
\end{equation}

\begin{equation}
\dot{A}_\mathrm{diss,i} = \frac{1}{2}\frac{\gamma-1}{\rho_i^{\gamma-1}}\frac{\dot{m}_{\mathrm{diss},i}}{m_i} \left[{\bf v}_i \cdot {\bf v}_i \right] - \frac{\dot{m}_{\mathrm{diss},i}}{m_i} A_i
\label{eqn:adotm}
\end{equation}
We will use the above correction terms throughout this paper. (We derive a general class of correction terms in Appendix \ref{sec:massdisscons}; these may lead to even better results in some situations, but we leave this as an investigation for future work.) It is clear that equations \ref{eqn:mdiss}, \ref{eqn:vdotm} and \ref{eqn:adotm} satisfy our criteria (iii) - (v) outlined above. For equal mass particle applications $\dot{m}_{\mathrm{diss},i} = 0$ by construction and equations \ref{eqn:mdiss}, \ref{eqn:vdotm} and \ref{eqn:adotm} have no effect.

A final concern is that adding mass dissipation will affect our solution of the continuity equation. Taking the time derivative of equation \ref{eqn:sphcont}, we have that: 

\begin{equation}
\dot{\rho}_i = \frac{d}{dt}\sum_j m_j W_{ij} = \sum_j \dot{m}_j W_{ij} + \sum_j m_j {\bf v}_{ij} \cdot \unabla W_{ij} 
\label{eqn:sphcontmassdiss}
\end{equation}
The right term is the familiar SPH continuity equation; the left term is a correction factor that accounts for mass dissipation. Thus, by using the familiar SPH density sum (equation \ref{eqn:sphcont}), we {\it automatically} include the mass dissipation correction to the continuity equation. However, we may still worry whether equation \ref{eqn:sphcontmassdiss} tends towards equation \ref{eqn:euler1} in the limit of infinite resolution. Substituting for $\dot{m}_j = \dot{m}_{\mathrm{diss},j}$ in equation \ref{eqn:sphcontmassdiss}, we have: 

\begin{equation}
\sum_j \dot{m}_j W_{ij} = \sum_{j,k} Q_{jk} (m_j - m_k) K_{jk} W_{ij} \simeq 0 
\end{equation} 
where $Q_{ij} = Q_{ji} = \overline{m}_{ij}/\overline{\rho}_{ij} \overline{\alpha}_{ij} v^p_{\mathrm{sig},ij}L_{ij}$, and the equation is very nearly vanishing since the sum is almost perfectly antisymmetric in the indices $j,k$ (the antisymmetry is broken by $W_{ij}$). In the continuum limit, the above term is exactly zero and so equation \ref{eqn:sphcontmassdiss} does indeed tend towards equation \ref{eqn:euler1} with increasing resolution. 

\begin{center}
\begin{figure*}
\hspace{3mm}\includegraphics[width=\textwidth]{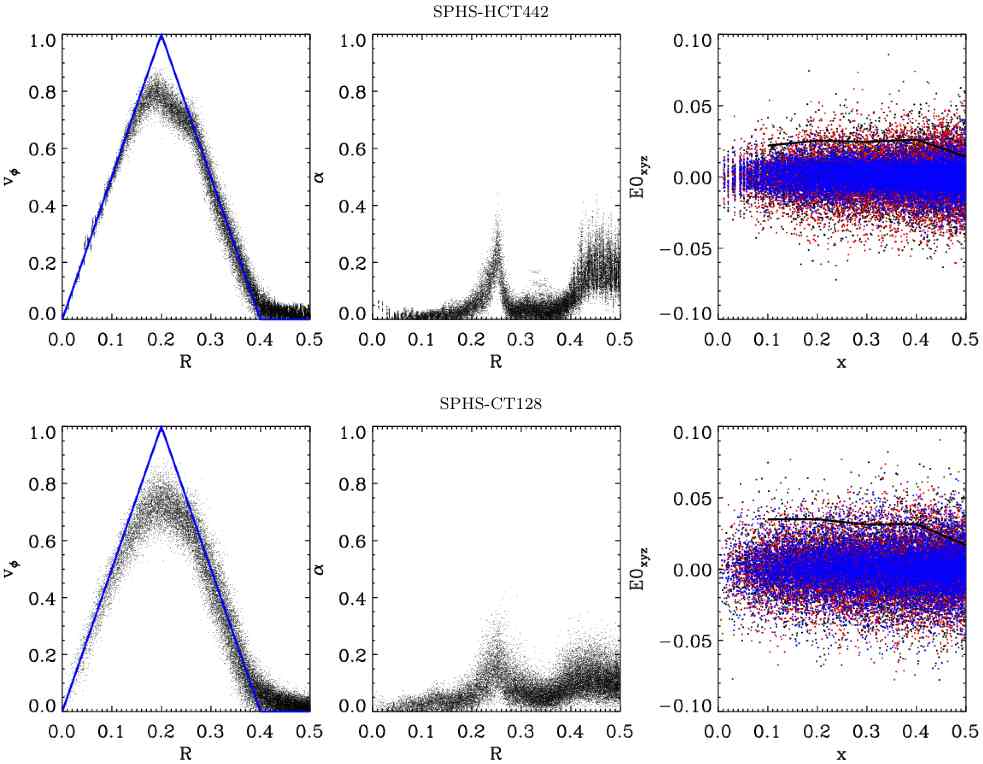}
\caption{Gresho vortex test results. The top panels show results for the HOCT4 kernel with 442 neighbours; the bottom panels for the CT kernel with 128 neighbours. The left panels show the $v_\phi$ velocities of particles after a time $t=1$ for $N=64\times64\times8$ particles; the analytic solution is marked in blue. The middle panels show the dissipation parameter $\alpha$ at the same time. The right panels show the three $\uE$ error components in black, red and blue. Over-plotted is the mean binned error magnitude $\muE$ (solid black line).} 
\label{fig:vortex}
\end{figure*}
\end{center}

\section{Timestepping}\label{sec:timestep} 

For our timestep control, we use individual particle timesteps ordered on a hierarchy of rungs in powers of two, as in \citet{2005MNRAS.364.1105S}. Particles are placed on rungs using a Courant-like condition: 

\begin{equation}
\Delta t_{i} = C \frac{h_i}{v_{\mathrm{max},i}}
\label{eqn:courant}
\end{equation}
where $C = 0.2$ is the Courant factor. We use this same fixed Courant factor for all tests presented in this paper. 

In addition to the standard timestep criteria above, we introduce a constraint similar to that in \citet{2009ApJ...697L..99S} to ensure that neighbouring particles do not differ in their timesteps by more than a factor of 4. 

\section{Implementation}\label{sec:implement}

We implemented our method in the {\tt GADGET-2} code \citep{2005MNRAS.364.1105S}. {\tt GADGET-2} is a massively parallel Tree-SPH code originally designed to model galaxy collisions, but now adapted for cosmological, hydrodynamic, magnetohydrodynamic and many other applications. \SPHG\ acts as a new hydro module within {\tt GADGET-2}, replacing the standard SPH parts of the code. We refer the reader to the original {\tt GADGET-2} paper for details of the gravity solver \citep{2005MNRAS.364.1105S}. 

The \SPHG\ hydro module, like {\tt GADGET-2} requires two loops over the particles. In the first loop, the densities are calculated (iterating to ensure equation \ref{eqn:fixedmass} is satisfied). At the same time, we calculate the polynomial fluid gradients (for the dissipation switch; \S\ref{sec:cross}). In the second loop, the hydrodynamic forces are evaluated along with the dissipation terms. Some speed comparisons between our current implementation of \SPHG\ and classic {\tt GADGET-2} SPH are given in \S\ref{sec:speed}.  

\section{Code tests}\label{sec:tests}

In this section, we present a suite of code tests designed to challenge the \SPHG\ method. In \S\ref{sec:sod}, we use the Sod shock tube test to examine shocks in \SPHG\ both with and without multimass particles, and to asses the rate of convergence in \SPHG. In \S\ref{sec:vortex}, we use the `Gresho' vortex test to examine convergence in shear flows in \SPHG, and the role of numerical viscosity. In \S\ref{sec:sedov}, we use a strong Sedov-Taylor blast wave test to see how well \SPHG\ performs in the presence of extreme entropy contrasts. In \S\ref{sec:kh}, we use a Kelvin-Helmholtz instability test with density contrast 1:8 -- both with and without multimass particles -- to examine mixing in \SPHG. Finally, in \S\ref{sec:blob}, we use the `blob' test -- a 1:10 density ratio gas sphere in a wind tunnel to assess how \SPHG\ performs in more complex flow situations where shocks and mixing combine. 
 
\subsection{Simulation labelling convention}\label{sec:label}

In the following subsections, we run a broad range of simulations both in our new hydrodynamics code \SPHG, and in `classic' SPH (the version of SPH that is in the public release version of the {\tt GADGET-2} code, and that is described in \bcite{2005MNRAS.364.1105S}). To avoid confusion, we use the following naming convention for these simulations: 

\begin{quote}
\begin{center}
SPHX-KKNNx
\end{center}
\end{quote} 
where X refers to the flavour of SPH: `classic' SPH (SPH), or our new code (\SPHG); KK refers to the kernel used (CS; cubic spline; equation \ref{eqn:cubicspline}), (CT; core-triangle; equation \ref{eqn:ctkern}), (HCT; High Order Core Triangle; equation \ref{eqn:hoctkern}); NN refers to the neighbour number (42, 96, 128, 442); and x is reserved to describe special simulations: x = g means that the test was run using glass rather than lattice initial conditions; x = e means that the test was run using a higher order momentum equation (equation \ref{eqn:limit} with the $\uE$ term subtracted); and x = multi means that the test was run using multimass particles. 

\subsection{Sod shock tube}\label{sec:sod} 

The Sod shock tube test is a 1D tube on the interval
$[-0.5,0.5]$ with a discontinuous change in properties at $x = 0$ designed
to generate a shock. The left state is described by $\rho_l = 1.0$,
$P_l = 1.0$, $v_l = 0$, and the right state by $\rho_r = 0.125$, $P_r
= 0.1$, $v_r = 0$, where $\rho, P$ and $v$ are the density, pressure
and velocity along the $x$ axis. We use an adiabatic equation of state
with $\gamma = 5/3$ and perform the test in 3D on the union
of a $32\times 32\times 400$ lattice on the left, with a $16\times 16\times 200$ lattice on
the right, giving a 1D resolution of $N_\mathrm{1D} = 600$ points. For the \SPHG\ simulation, we set an initial dissipation parameter $\alpha = 1$ over the initial pressure discontinuity ($-0.05 < x < 0.05$) since this test starts with a shock. We use lattice ICs for this test aligned with the shock, identical to those presented in \Ron. In SPH, however, there is some freedom in how to lay down the particles. \citet{2010MNRAS.408..669C}, for example, use instead glass-like initial conditions. These are noisier than the simple cubic lattice we use here, but have no preferred direction. While the choice of initial condition can affect the results, it should not affect the difference in the results between SPH and \SPHG. We will demonstrate this using glass and lattice ICs for the Gresho vortex test in \S\ref{sec:vortex}.

The results at time $t = 0.2$ are shown in Figure \ref{fig:sod} for SPH (red) and \SPHG\ (black). The analytic solution is marked in blue. Notice that \SPHG\ performs well on this test. In particular, the pressure blip at the shock, present in the SPH run (red), is almost completely gone. The SPH run, which used 96 neighbours, shows significantly more noise in the velocity distribution. This occurs due to symmetry breaking of the simple cubic lattice ICs and is reduced for glass-like initial conditions \citep{2010MNRAS.408..669C}. It is also reduced by moving to higher order kernels that have larger neighbour number and are therefore correspondingly less noisy \citep[e.g.][]{2010arXiv1012.1885P}. This is why the noise is not present in the \SPHG\ simulation.

In addition, we perform this same test using multimass particles in \SPHG, where we sample the domain uniformly with a $16\times16\times400$ lattice -- the same resolution as the low density phase in the single particle mass Sod test (Figure \ref{fig:sod}, bottom panels). This is an extremely challenging test for \SPHG. The initial conditions have a sharp jump in three fluid variables: entropy, pressure and mass. Nonetheless, the solution is in excellent agreement with the analytic result.

Finally, we perform a convergence study for the Sod shock tube test in Figure \ref{fig:sod_converge}. We follow \citet{2010ARA&A..48..391S} and define our error measure as: 

\begin{equation}
L1(v_x) = \frac{1}{N_b}\sum_i^{N_b} |\overline{v}_{x,i} - \overline{v}_{x}(x_i)|
\end{equation}
where $N_b$ is the number of bins in $x$, $\overline{v}_{x,i}$ is the mean x-velocity in bin $i$, and $\overline{v}_{x}(x_i)$ is the expected analytic mean velocity in bin $i$. We use a $x$-bin width of 0.01. This is small enough to capture the improvements with increasing resolution, but not so small as to over-sample the lowest resolution run. 

\citet{2010ARA&A..48..391S} find that for shock tests in 2D, SPH performs worse than the optimal $N_{1D}^{-1}$ scaling, giving something closer to $N_{1D}^{-0.7}$. Here we find that, by contrast, \SPHG\ gives a near-optimal scaling, going as very nearly $N_{1D}^{-1}$, except at the highest resolution (compare the black and thick black lines in Figure \ref{fig:sod_converge}). The slowing down of the convergence rate for the highest resolution simulation is due to the fundamental convergence limit set by our neighbour number. We demonstrate this in more detail for the Gresho Vortex test, next. 

\subsection{Gresho Vortex test}\label{sec:vortex}

We set up the Gresho Vortex test similarly to \citet{2010ARA&A..48..391S} \citep[and see][]{gresho}. The test involves an $N\times N\times N/8$ 3D lattice of particles. A velocity and pressure field are applied to these to set up a stable vortex: 

\begin{equation}
v_\phi(R) = \left\{
\begin{array}{lll}
5R & \mathrm{for} & 0 \le R \le 0.2 \\
2 - 5R  & \mathrm{for} & 0.2 \le R \le 0.4 \\
0 & \mathrm{for} & R \ge 0.4 \\
\end{array}
\right.
\end{equation}

\begin{equation}
P(R) = \left\{
\begin{array}{lll}
5 + \frac{25}{2} R^2 & \mathrm{for} & 0 \le R \le 0.2 \\
9 + \frac{25}{2} R^2 - & & \\
20 R + 4 \ln(R/0.2) & \mathrm{for} & 0.2 \le R \le 0.4 \\
3 + 4 \ln 2 & \mathrm{for} & R \ge 0.4 \\
\end{array}
\right.
\end{equation}
where $R = \sqrt{x^2+y^2}$ and we set $\rho = 1$ and $\gamma = 5/3$. 

The above vortex should be stable over many rotations, but in practice will decay due to the numerical viscosity inherent in the scheme. As such, it is a useful test of the numerical viscosity generated in shear flows. Indeed, classic SPH performs poorly on this test converging very slowly to the wrong solution \citep{2010ARA&A..48..391S}. Such rotating configurations are common in a wide class of astrophysical problems; it is important for numerical schemes to perform well on such tests. 

The results for \SPHG\ are given in Figure \ref{fig:vortex}. We show, from left to right, the rotational velocity profile of the vortex (black points) as compared to the analytic solution (thick blue line); the dissipation parameter $\alpha$, and the leading order error in the momentum equation $\uE$. We find significantly better performance in \SPHG\ than was found by \citet{2010ARA&A..48..391S} for SPH. The primary reason for this -- surprisingly -- is not the lower viscosity of the method. The average viscosity {\it is} lower in \SPHG -- in the range $0.05 < \alpha < 0.3$ as compared to SPH that has constant $\alpha = 1$ (see Figure \ref{fig:vortex}, middle panels). But, the real reason for the improvement is the improved {\it force accuracy}. In Figure \ref{fig:vortex}, the top three panels show the results for our default method (SPHS-HCT442), while the bottom three show results for SPHS using a lower neighbour number with the CT kernel (SPHS-CT128; see \S\ref{sec:label} for our simulation labelling convention). If anything, the viscosity is slightly {\it lower} for the SPHS-CT128 simulation, yet the results are worse, with increased noise and a bias in the rotational velocity for $R \simlt 0.2$. The only difference between these two simulations is the neighbour number, and the associated $\uE$ error. Indeed, in SPHS-HCT442, the $\uE$ is lower than for the SPHS-CT128 simulation (see Figure \ref{fig:vortex}, right panels). 

\begin{center}
\begin{figure}
\begin{center}
\hspace{-6mm}
\includegraphics[width=0.49\textwidth]{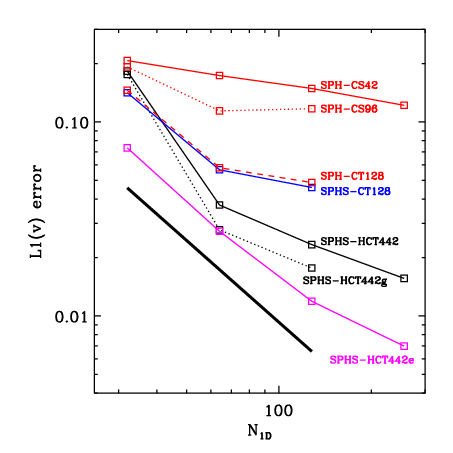}
\caption{Gresho vortex convergence test results. The x-axis gives the number of particles along one side of the box $N_{1D}$ (Figure \ref{fig:vortex} shows results for $N_{1D}=64$); the y-axis shows the binned velocity error in $R$-bins of width 0.01. The different line colours show different SPH methods; the naming convention (marked) is as described in \S\ref{sec:label}. Three simulations: SPHS-CS42 (`classic' SPH); SPHS-HCT442 (our default \SPHG\ method); and SPHS-HCT442e (our default method using a higher order momentum equation) are tested to higher resolution ($N_{1D} = 256$). The thick black line shows a scaling of $N_{1D}^{-1.4}$ (the best-possible scaling is $N_{1D}^{-2}$).}
\label{fig:vortex_converge}
\end{center}
\end{figure}
\end{center}

\begin{center}
\begin{figure*}
\hspace{-11mm}
\includegraphics[width=\textwidth]{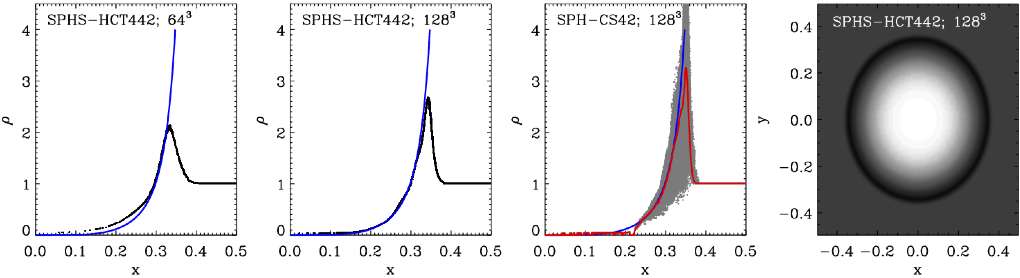}\hspace{-8mm}
\caption{Sedov-Taylor blast wave test results. {\bf Left three panels:} The density profile of the gas at $t=0.05$ for $N=64^3$ and $N=128^3$ particles for SPHS-HCT442, and for $N=128^3$ particles for SPH-CS42 (using the timestep limiter described in \S\ref{sec:timestep}). The blue lines mark the analytic solution. For the SPHS simulations, the actual un-binned point particle densities are plotted in black; for the SPH simulation, they are plotted in grey. Notice the significantly larger noise for the SPH simulation. A mean binned profile, using a bin size of $\Delta x = 0.001$ is over-plotted in red. {\bf Right-most panel:} Logarithmic density contours of the blast wave viewed from top at $t=0.05$ for SPHS-HCT442 with $N=128^3$ particles.}
\label{fig:sedov}
\end{figure*}
\end{center}
 
In Figure \ref{fig:vortex_converge}, we explore this further by presenting convergence tests for SPH and SPHS using varying neighbour number and kernel choice. We calculate the $L1(v_\phi)$ error norm as in the Sod test (\S\ref{sec:sod}), using a bin size of $\Delta R = 0.01$. The thick black line on the plot marks the ideal scaling of $N_{1D}^{-2}$ (ideal for a second order method away from contact discontinuities). The red lines show results for classic SPH with 42 neighbours (solid line), 96 neighbours (dotted line) and 128 neighbours using the CT kernel (dashed line). Notice that SPH-CS42 converges very slowly with increasing resolution, with the error always larger than 10\%. Increasing the neighbour number to 96 neighbours helps at low resolution but gives diminishing returns with increasing resolution. This agrees with our results from \Ron, where we showed that in shear flows the $\uE$ error improves only very slowly with increasing neighbour number for the CS kernel. This is because for neighbour number larger than $\sim 40$, the particles begin to clump preventing any significant improvement in the kernel sampling. By contrast, switching to the CT kernel with 128 neighbours -- that is manifestly stable to particle clumping -- gives a significant improvement in the convergence rate (dashed line). Now the error drops to $\sim 5$\% for $N_{1D} = 128$. The solid blue line shows the result for SPHS using the CT kernel with 128 neighbours (SPHS-CT128). The results are only very slightly better than for classic SPH. This highlights that, for this test, it is the {\it force accuracy} that determines the rate of convergence, not the dissipation scheme. The solid black line shows the result for our default SPHS scheme: SPHS-HCT442. With 442 neighbours and a correspondingly higher order kernel, our method now converges on percent level accuracy for this test. The convergence appears to be uninterrupted even at $N_{1D} = 256$, though the rate is perhaps slowing. The dotted black line shows the results for the same simulation but run using glass initial conditions. The error is slightly improved, but the rate of convergence is identical. This demonstrates that our results are not sensitive to the initial particle distribution. Finally, the solid magenta line shows results for our default SPHS scheme, but using a higher order momentum equation. For this simulation, we replaced equation \ref{eqn:sphmoment}, with the left term in equation \ref{eqn:limit} -- i.e. explicitly subtracting away the $\uE$ error (SPHS-HCT442e). In this case, we should expect to see a steady convergence rate, without any need to further increase the neighbour number. Indeed, this is what is seen. The results are now significantly better than any of the other SPH methods. At the very highest resolution ($N_{1D} = 256$), however, there may still be some slowing in the convergence rate. It is possible that this is simply a fluctuation (running the test at even higher resolution to test this seems extravagant). Alternatively, it may be that at these resolutions the viscosity does start to play a role, slowing the convergence rate.

The most promising results for this particular test come from the SPHS-HCT442e method that uses a higher order momentum equation. Unfortunately, this same equation violates pairwise momentum conservation between particles and causes problems in strong shocks (\S\ref{sec:errors}; and see \bcite{1996PASA...13...97M}). As such, we defer the investigation of such schemes to future work. We note here, however, that our default scheme -- SPHS-HCT442 -- still performs very well on this test, achieving percent level accuracy with increasing resolution. As we have shown already, this is also true for the Sod test (\S\ref{sec:sod}; and see Figure \ref{fig:sod_converge}). This suggests that for most astrophysics applications of interest, where the errors are in any case dominated by sub-grid physics prescriptions rather than the hydrodynamics solver, our default scheme -- SPHS-HCT442 -- should be sufficient, without any need to further raise the neighbour number. 

\subsection{Sedov-Taylor blast wave}\label{sec:sedov} 

We set up a Sedov-Taylor blast-wave test as in \citet{2002MNRAS.333..649S} using a uniform lattice of $64^3$ particles with initial density $\rho = 1$. We inject an explosion energy $E=1$ into a central region $r<0.08$. This corresponds to an initial entropy per central particle of $A = 350$. The remaining particles are assigned $A = 0.05$, giving an entropy contrast of $\sim 7000$. The analytic similarity solution to this problem is well known \citep[see e.g.][]{1966hydr.book.....L}, and gives a time evolution for the blast wave radius of:

\begin{equation}
r(t) = 1.15 \left(\frac{E t^2}{\rho}\right)^{1/5}
\end{equation}
for an adiabatic index of $\gamma = 5/3$. 

The Sedov-Taylor test is particularly challenging for any hydrodynamical code because of the extreme entropy gradient in the initial conditions. The results for \SPHG\ for $N=64^3$ and $N=128^3$ particles are given in the left two panels of Figure \ref{fig:sedov}. As the resolution is increased, the results converge on the analytic solution shown in blue: the peak density of the shock increases, while the low density tail better matches the analytic expectations. The blast wave is perfectly symmetric, as shown in the right-most panel. For comparison, the results for classic SPH (with $N = 128^3$) are shown in the third panel. Notice that the result is significantly more noisy (compare the grey dots with the black dots in the left two panels). The reduced noise in \SPHG\ is partly due to the increased neighbour number, and partly due to the entropy dissipation (see for example similarly less noisy results for this test reported in \bcite{2007MNRAS.379..915R}). The mean solution for SPH is, however, in good agreement with the analytic solution (compare the red and blue lines for the SPH-CS42 panel). Note that, for this test we had to use the timestep limiter described in \S\ref{sec:timestep} (and see \bcite{2009ApJ...697L..99S}), and so this simulation is not strictly speaking `classic' SPH as we have defined it in this paper. Similar results can be obtained with classic SPH by using fixed timesteps and a sufficiently small Courant factor (equation \ref{eqn:courant}). However, this is computationally very expensive. 

One interesting aspect of the Sedov test is that it allows us to compare the spatial resolution in SPHS-HCT442 with classic SPH using 42 neighbours (SPH-CS42). Notice that the SPH-CS42 simulation resolves higher density in the shock. The unbinned particles reach densities up to $\rho_\mathrm{max} \sim 4.5$ in simulation units, whereas our default scheme (SPH-HCT442) manages only $\rho_\mathrm{max} = 2.7$. We argued in \S\ref{sec:errors} that the HOCT4 kernel with 442 neighbours should degrade the spatial resolution by a factor $f \sim 1.5$ as compared to the CS kernel with 42 neighbours. Since the shock front for the Sedov test is one dimensional, then we can expect a lower peak density in SPHS-HCT442 of a factor $\sim f$. This is almost exactly what is seen since $\rho_\mathrm{max,SPH} / \rho_\mathrm{max,SPHS}  = 1.67$. In practice, however, the spatial resolution of the SPH simulation is not this good because of the increased noise. We ought to trust only the averaged solution that is shown in red. For this averaged density, the peak is significantly lower, with $\rho_\mathrm{max} \sim 3.3$ -- only a factor 1.2 better than our default SPHS method. We conclude from this that SPHS-HCT442 does not significantly degrade the spatial resolution as compared to SPH-CS42, especially once the reduced noise of the method is taken into account.

\begin{center}
\begin{figure*}
\hspace{-6mm}\includegraphics[width=\textwidth]{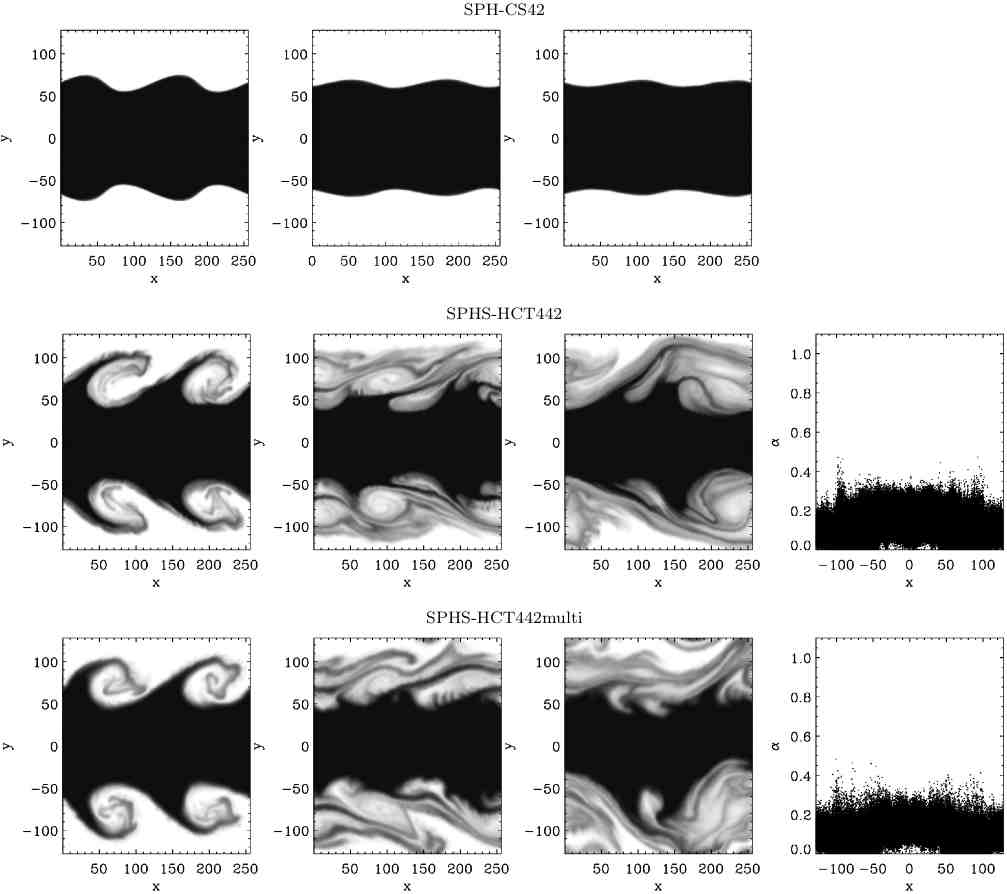}
\caption{KH1:8 test results. From left to right panels show: logarithmic density contours at: $\tau_\mathrm{KH}=1, 2$ and $3$; and the dissipation parameter $\alpha$ at $\tau_\mathrm{KH}=1$. The top panels show results for a single particle mass simulation ($N=2,359,296$); the bottom panels for a multimass simulation with a uniform density particle distribution ($N=524,288$).}
\label{fig:kh}
\end{figure*}
\end{center}

\subsection{Kelvin-Helmholtz test}\label{sec:kh} 

We set up a 1:8 density contrast Kelvin-Helmholtz (KH) test in 3D as in \Ron. We used a
periodic thin slab defined by $x\in \{-0.5,0.5\}$, $y\in \{-0.5,0.5\}$
and $z\in \{-1/64,1/64\}$. The domain satisfied: 
\begin{equation}
\rho,T,v_x=\left\{
 \begin{array}{rl} 
 	\rho_1,T_1,v_1  & |y|< 0.25\\
 	\rho_2,T_2,v_2  & |y|> 0.25
 \end{array} \right.
\end{equation}
The density and temperature ratio were $R_\rho=\rho_{\rm 1}/\rho_{\rm
  2}=T_{\rm 2}/T_{\rm 1}=c_{\rm 2}^2/c_{\rm 1}^2$, ensuring that the
whole system was in pressure equilibrium. The two layers were given
constant and opposing shearing velocities, with the low density layer
moving at a Mach number $\mathcal{M}_{\rm{2}}=-v_{\rm 2}/c_{\rm 2}
\approx 0.11$ and the dense layer moving at
$\mathcal{M}_{\rm{1}}=\mathcal{M}_{\rm{2}}\sqrt{R_\rho}$. The density
ratios considered in this work are small which assures a subsonic
regime where the growth of instabilities can be treated using equation
\ref{eq:KHI} \citep{1997ApJ...483..262V}.
 
To trigger instabilities, velocity perturbations were imposed on the
two boundaries of the form:
\begin{eqnarray}
v_y & =& \delta v_{\rm y}[\sin(2\pi
(x+\lambda/2)/\lambda)\exp(-(10(y-0.25))^2) \nonumber \\
& & - \sin(2\pi x/\lambda)\exp(-(10(y+0.25))^2)]
\end{eqnarray}
where the perturbation velocity $\delta v_{\rm y}/v=1/8$ and
$\lambda=0.5$ is the wavelength of the mode.

The linear growth rate of the KHI is given by:
(\bcite{1961hhs..book.....C}):
\begin{equation}
w=k\frac{(\rho_1\rho_2)^{1/2}v}{(\rho_1+\rho_2)},
\end{equation} 
where $k=2\pi/\lambda$ is the wavenumber of the instability, $\rho_1$
and $\rho_2$ are the densities of the respective layers and $v=v_1-v_2$ is
the relative shear velocity. The characteristic growth time for the
KHI is then: 
\begin{equation}
\label{eq:KHI}
\tau_{\rm
  KH}\equiv\frac{2\pi}{w}=\frac{(\rho_1+\rho_2)\lambda}{(\rho_1\rho_2)^{1/2}v}.
\end{equation}

We set up two simulations to satisfy the setup described above. An equal mass particle simulation with $N=2,359,296$, and a multimass version with $N=524,288$. The latter simulation used a {\it uniform grid} of particles, with mass ratio 1:8 to describe the density step. To satisfy pressure equilibrium everywhere, the temperatures were adjusted at the boundary to be consistent with the SPH density step that is smooth (c.f. equation \ref{eqn:sphcont}). 
 
The results of the test at times $\tau_{\rm KH} = 1,2$ and $3$ are shown in Figure \ref{fig:kh}. The top row shows the results for classic SPH (SPH-CS42); the middle row shows the results for our default SPHS scheme (SPHS-HCT442), using single mass particles; the bottom row shows the results for SPHS-HCT442 using multimass particles. The SPH results are poor, with no mixing observed between the fluid layers, similarly to what has been reported in previous works (e.g. \bcite{2006astro.ph.10051A}; \Ron). By contrast, the SPHS results show the growth of KH rolls on the correct timescale and resolved mixing into the fully non-linear regime. Furthermore, the single mass simulation and multimass simulation (middle and bottom panels) are in excellent agreement. They differ in the details of the non-linear evolution caused by the growth of smaller noise-seeded rolls. But the growth time for the primary KH roll is in excellent agreement with analytic expectations, while the non-linear evolution is qualitatively similar. The multimass simulation is slightly more diffusive due to the additional mass dissipation between particles at the boundary. However, this simulation (because of the lower particle number) ran almost 5 times faster. 

As discussed in our previous paper (\Ron), the improved performance for the KH test in \SPHG\ is a result of both the improved force accuracy (due to the increased neighbour number and higher order, stable, kernel), {\it and} the improved dissipation. We demonstrate this in Appendix \ref{sec:khdiss}, where we show the effect of switching off the entropy and mass dissipation terms (equations \ref{eqn:entdiss} and \ref{eqn:mdiss}) for this test. 

\begin{center}
\begin{figure*}
\hspace{2mm}\includegraphics[width=\textwidth]{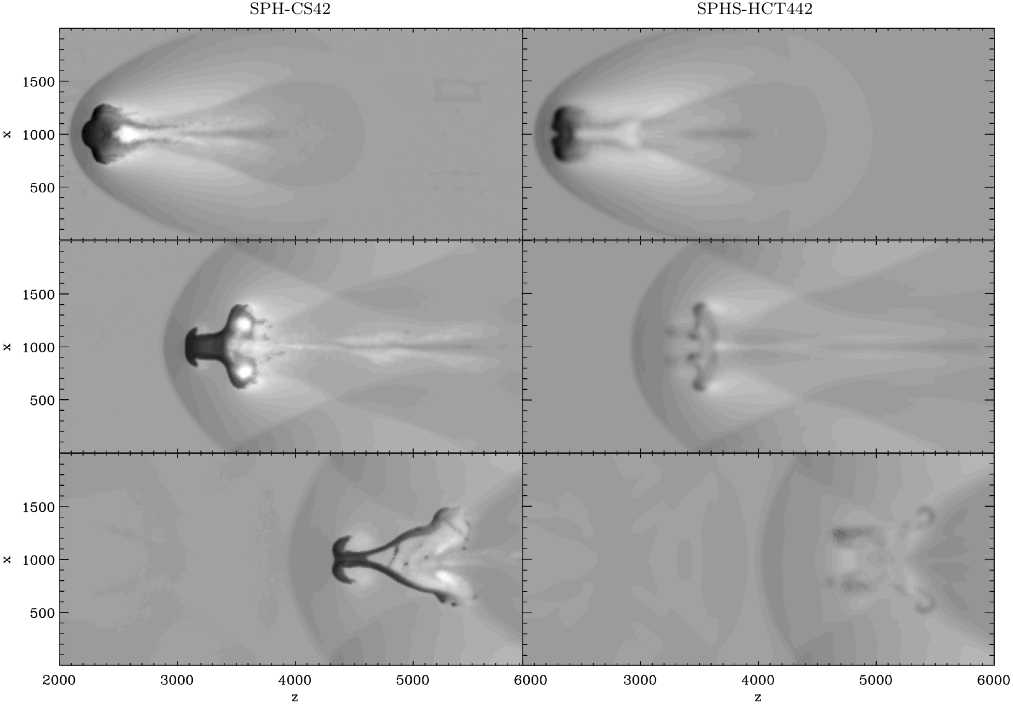}
\caption{Blob test results at $\tau_\mathrm{KH}=1$ (top), $2$ (middle), and $3$ (bottom) for classic SPH (left) and SPHS (right). All plots show logarithmic density contours.}
\label{fig:blob}
\end{figure*}
\end{center}

\subsection{The `blob' test}\label{sec:blob} 

The `blob' test is a spherical cloud of gas of
radius $R_{\rm cl}$ in a wind 
tunnel with periodic boundary conditions. The ambient medium is ten 
times hotter and ten times less dense than the cloud so that the system is in
pressure equilibrium. The wind velocity ($v_{\rm wind}=c_s\mathcal{M}$) has an
associated Mach number $\mathcal{M}=2.7$. This leads to the formation
of a bow shock after which the post-shock subsonic flow interacts with
the cloud and turns supersonic as it flows past it. The test was first presented (with a full analytic analysis) in \citet{2006astro.ph.10051A}. Here, we set up the test as in \Ron\ with $N=126,744$ in the blob, arranged on a lattice. As in \Ron, we seed an initial inwards perturbation on the blob surface. 
 
The results at times $\tau_\mathrm{KH}=1$ (top), $2$ (middle), and $3$ (bottom) in classic SPH (left) and SPHS (right) are given in Figure \ref{fig:blob}. In classic SPH, similarly to what has been reported in previous works, the blob does not break up and survives for the full length of the simulation. Furthermore, the suppression of instabilities at the fluid interface is sufficient to remove the inward perturbation that was seeded in the initial conditions. By contrast, this perturbation is clearly visible in the SPHS simulation and grows causing the blob to split down the middle in good agreement with both Eulerian codes and our OSPH method (\Ron). Finally, notice that the symmetry of the blob is well-preserved even at $\tau_\mathrm{KH}=3$ -- well into the non-linear regime. 

\begin{center}
\begin{figure*}
\hspace{3mm}\includegraphics[width=\textwidth]{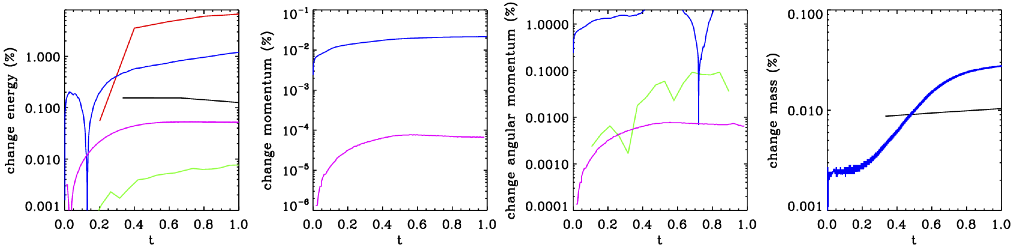}
\caption{Conservation in \SPHG. From left to right, the panels show conservation of energy, momentum, angular momentum, and (where relevant) mass for the simulation test suite presented in this paper. The coloured lines show results for: the multimass Sod test (black); the Sedov-Taylor test (red; for $N=128^3$ particles); the Gresho Vortex test (green; for $N=64\times64\times8$); the multimass KH1:8 test (blue); and the Blob test (purple). The results are normalised to a simulation time of 1, where ``1" is the maximum time presented in this paper (i.e. for the KH1:8 test this is $\tau_\mathrm{KH} = 3$). Momentum and angular momentum conservation results are only shown where these quantities are not zero in the initial conditions (to avoid a divide by zero in the percentage errors).}
\label{fig:conserve}
\end{figure*}
\end{center}

\subsection{Conservation}\label{sec:conservation}

Figure \ref{fig:conserve} summarises the conservation performance of \SPHG\ for all of our tests. From left to right, we show the conservation of energy, momentum, angular momentum and (where relevant) mass. The results are normalised to a simulation time of 1, where ``1" is the maximum time presented in this paper (i.e. for the KH1:8 test this is $\tau_\mathrm{KH} = 3$). Momentum and angular momentum conservation results are only shown where these quantities are not zero in the initial conditions (to avoid a divide by zero in the percentage errors). The worst performance is for the Sedov-Taylor test that conserves energy at the $\sim 5$\% level. However all other tests conserve energy, mass, momentum and angular momentum to better than 1\% over the full simulation time.
 
\subsection{Code performance}\label{sec:speed}  

\begin{center}
\begin{figure}
\hspace{30mm}SPHS-HCT442 $\mid$ {\color{red} SPH-CS442}\vspace{-3mm}\\
\includegraphics[height=0.49\textwidth]{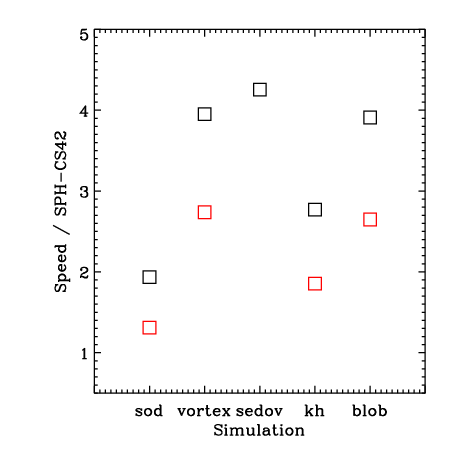}
\caption{The ratio of the speed of our default \SPHG\ scheme (SPHS-HCT442; black squares) and SPH-CS442 (red squares) to `classic' SPH (SPH-CS42). (We use the simulation naming convention as described in \S\ref{sec:label}.) For all tests, we used 16 processors. The Sod tests were compared at the $N_\mathrm{1D}=600$ resolution; the vortex tests at $N_\mathrm{1D} = 128$; and the Sedov tests at $N_\mathrm{1D} = 128$. In all cases, we eliminate the start-up time costs (the time taken to complete step zero).}
\label{fig:timings_plot}
\end{figure}
\end{center}

Figure \ref{fig:timings_plot} compares the ratio of the speed of our default \SPHG\ scheme (SPHS-HCT442; black squares) and SPH-CS442 (red squares) to classic SPH (SPH-CS42). (We use the simulation naming convention as described in \S\ref{sec:label}.) For all tests, we used 16 processors. The Sod tests were compared at the $N_\mathrm{1D}=600$ resolution; the vortex tests at $N_\mathrm{1D} = 128$; and the Sedov tests at $N_\mathrm{1D} = 128$. In all cases, we eliminate the start-up time costs (the time taken to complete step zero). There is some significant variation in speed across all of the tests with the cost of SPHS ranging from 2 to 4 times that of SPH-CS42, but typically SPHS is 3-4 times slower at like particle number. 

Note that the above speed tests are conservative. We could equally well conduct the tests at like {\it numerical error}, rather than like particle number. For the Gresho vortex test, for example, it is unlikely that SPH-CS42 can ever achieve equivalent accuracy to SPHS for any reasonable particle number (see Figure \ref{fig:vortex_converge}). To obtain $\sim 1$\% accuracy on this test, classic SPH would require an enormous particle number and be significantly slower than SPHS. 

Finally, we have not made any attempt to optimise our current implementation of \SPHG. Faster neighbour search algorithms, or neighbour caching could conceivably gain back much of the speed losses as compared to classic SPH. In addition, for real astrophysics applications, the additional work done on the neighbours may be compensated by improved timestepping (due to the reduced noise), and better load balancing in highly parallel simulations. Such considerations are beyond the scope of this present work. 

\section{Conclusions}\label{sec:conclusions}

We have presented an implementation of Smoothed Particle Hydrodynamics (\SPHG) that has two novel features. The first is an improved treatment of dissipation. We use the spatial derivative of the velocity divergence as a higher order dissipation switch. Our switch -- which is second order accurate -- detects flow convergence {\it before} it occurs. If particle trajectories are going to cross, we switch on the usual SPH artificial viscosity, as well as conservative dissipation in all advected fluid quantities (for example, the entropy). The viscosity and dissipation terms (that are numerical errors) are designed to ensure that all fluid quantities remain single-valued as particles approach one another, to respect conservation laws, and to vanish on a given physical scale as the resolution is increased. The second novel feature is the use of significantly larger neighbour number (442) to improve the force accuracy. As in our previous work, we use a novel kernel function that is: (i) higher order such that the spatial resolution is not significantly degraded by our larger neighbour number; and (ii) that has a constant first derivative in the centre to prevent particle clumping (this latter ensures a smooth particle distribution on the kernel scale, which is necessary to obtain the improvement in the force accuracy).  

We have demonstrated that \SPHG\ alleviates a number of known problems with `classic' SPH\footnote{We define `classic' SPH as that implemented in the public release version of the {\tt GADGET-2} code (\bcite{2005MNRAS.364.1105S}).}, successfully resolving mixing, and recovering numerical convergence with increasing resolution. An additional key advantage is that -- treating the particle mass similarly to the entropy -- we are able to use multimass particles, giving significantly improved control over the refinement strategy. 

We have presented a wide range of code tests: the Sod shock tube, Sedov-Taylor blast wave, Gresho vortex, Kelvin-Helmholtz instability, the `blob test', and some convergence tests. Our method performed well on all tests, giving good agreement with analytic expectations. For some tests, like the Gresho vortex, most of the improvement over `classic' SPH is due to the improved force accuracy. For other tests like the (high density contrast) Kelvin-Helmholtz instability, the improved dissipation is equally important. We deliberately picked challenging tests that involve sharp features in one or more of the fluid quantities. These are inherently difficult to resolve for our method that is manifestly smooth, yet we show that \SPHG\ copes well even in such situations. 

In our current implementation (that is likely sub-optimal) \SPHG\ is typically 3-4 times slower than `classic' SPH (using 42 nearest neighbours) for like particle number. However, this additional cost should be offset against the improvement in the quality of the hydrodynamic solution in SPHS, the significantly reduced noise, and the improved rate of convergence. For the Gresho vortex test, for example, SPHS achieves $\sim$ percent level accuracy as compared to $\sim 10$\% in SPH for the same particle number. 

The main remaining flaw in the SPHS algorithm is its low order. This means that formal convergence requires the neighbour number to be increased along with the particle number (using increasingly higher order stable kernels to maintain spatial resolution). However, our default kernel choice with 442 neighbours is already sufficient to obtain percent level accuracy on the hydrodynamic tests we present here. It is unlikely that the neighbour number will need to be increased further than this for most astrophysical applications of interest.
 
\SPHG\ will be useful for any astrophysics application involving multiphase fluid flow (e.g. resolving the ISM in galaxy discs), or where the use of multimass particles would be advantageous. We will apply it to a broad range of problems in forthcoming papers.

\section{Acknowledgements}

JIR would like to acknowledge support from SNF grant PP00P2\_128540/1. This research used the ALICE High Performance Computing (HPC) facility at the University of Leicester and the Brutus HPC facility at the ETH Z\"urich. We would like to thank Peter Creasey for kindly supplying his Sedov-Taylor test initial conditions and solution, and for pointing out that OSPH does not perform well for the Sedov-Taylor test. We would like to thank Walter Dehnen, Lee Cullen, Chris Nixon, Alex Hobbs and Romain Teyssier for useful discussions. We would like to thank the referee for useful comments that have improved the clarity of this work. Finally, we would like to thank Volker Springel for making his {\tt GADGET-2} code available to the community, and for useful discussions. 

\appendix

\section{A fully conservative version of SPHS}\label{sec:fullcons}

\begin{center}
\begin{figure}
\hspace{-5mm}
\includegraphics[height=0.45\textwidth]{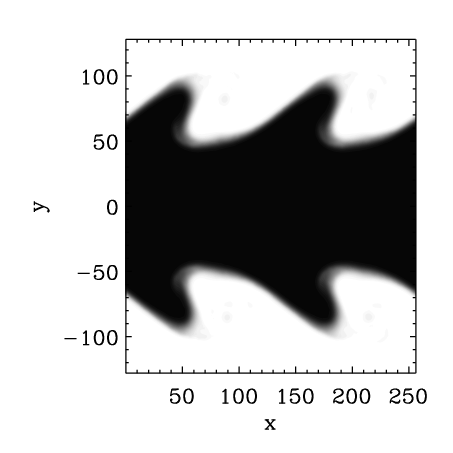}
\caption{KH1:8 multimass test results at $\tau_\mathrm{KH}=1$ for a fully conservative version of \SPHG\ (that uses equation 
\ref{eqn:sphmomentcons} instead of equation \ref{eqn:sphmoment}). Notice that the results are significantly more diffusive than our default scheme shown in Figure \ref{fig:kh}.}
\label{fig:fullcons}
\end{figure}
\end{center}

A fully conservative version of \SPHG\ can be constructed by replacing equation \ref{eqn:sphmoment} with that in \citet{2002MNRAS.333..649S}: 

\begin{equation}
\frac{d\uv_i}{dt} = -\sum_j^N m_j \left[f_i \frac{P_i}{\rho_i^2} \nabla_i W_{ij}(h_i) + f_j \frac{P_j}{\rho_j^2} \unabla_i W_{ij}(h_j)\right]
\label{eqn:sphmomentcons}
\end{equation}
where the function $f_i$ is a correction factor that ensures energy conservation for varying smoothing lengths:
\begin{equation}
f_i = \left(1 + \frac{h_i}{3\rho_i}\frac{\partial \rho_i}{\partial h_i}\right)^{-1};
\end{equation}
As discussed in \Ron, the above momentum equation gives improved (in fact manifest) energy conservation, but larger truncation error. For applications where energy conservation is of paramount importance (for example, where a system is evolved for many dynamical times), the above equation should be used. However, in this case, care must also be taken over the timestepping \citep[e.g.][]{2011EPJP..126...55D}. For the tests presented in this paper, the energy losses due to variable timesteps dominate and the above momentum equation gains only $\sim 0.5$\% in energy conservation for the KH1:8 multimass test (see \S\ref{sec:conservation}); and $\sim 2$\% for the Sedov test (a factor $\sim 2$ improvement in both cases). However, the larger truncation error introduces significantly more diffusion. This is shown in Figure \ref{fig:fullcons}, where we present results for the multimass KH1:8 test (see \S\ref{sec:kh}) at time $\tau_\mathrm{KH} = 1$, using equation \ref{eqn:sphmomentcons}. For this reason, our default choice for \SPHG\ is the momentum equation \ref{eqn:sphmoment}. 

\section{The trouble with `RT' pressures}\label{sec:pestimate}

\begin{center}
\begin{figure}
\begin{center}
\hspace{-8mm}\includegraphics[height=0.45\textwidth]{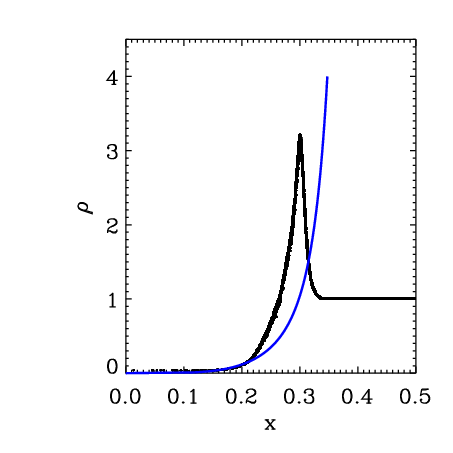}
\caption{Sedov-Taylor blast wave test results using the `RT' pressure estimator (equation \ref{eqn:rtp}) without entropy dissipation. The plot shows the density profile of the gas at time $t=0.05$, similarly to Figure \ref{fig:sedov}. Notice that the shock front is displaced with respect to the analytic curve (blue).}
\label{fig:sedovrt}
\end{center}
\end{figure}
\end{center}

In \Ron, we used the same equations of motion as \SPHG, but with the pressure estimator in equation \ref{eqn:rtp}. This ensured manifestly smooth pressures throughout the flow, allowing us to successfully model mixing between different fluid phases. However, while equation \ref{eqn:rtp} gives excellent performance for multiphase flow applications, it performs poorly in strong shocks where the entropy gradients on the kernel scale are large. We show this in Figure \ref{fig:sedovrt}, where we plot results for the Sedov-Taylor blast wave problem (with $N=128^3$; and see \S\ref{sec:sedov}), using the `RT' pressure estimator (equation \ref{eqn:rtp}) without entropy dissipation. As can be seen, the resulting shock front, while very smooth, is not in good agreement with the analytic curve shown in blue. 

The above highlights the key problem with `RT' densities and pressures. Particles in the kernel with very different entropies are down-weighted in the sum. This means that to obtain good kernel sampling, we must scale the neighbour number with the entropy contrast on the kernel scale. For the Sedov-Taylor blast wave, where the initial entropy contrast is $\sim 7000$, this is prohibitively expensive. Not doing this, however, leads to a significant numerical error as can be seen in Figure \ref{fig:sedovrt}. For this reason, in this paper, we have abandoned the density and pressure estimators given in equations \ref{eqn:sphcontrtent} and \ref{eqn:rtp}. Instead, we ensure smooth pressures through our higher order dissipation switch described in \S\ref{sec:cross}. 

\section{Fitting an $N$th order polynomial to a fluid quantity}\label{sec:full} 

We describe here an algorithm for fitting an order $N$ polynomial to an irregular point distribution \citep[see e.g.][]{fan+gijbels96,2003ApJ...595..564M}. We give the relevant equations for a second order fit in three dimensions, but the method straightforwardly generalises to arbitrary order and dimension. Assuming that a fluid quantity $q_i$ defined at particle position $i$ is smooth (and therefore differentiable), we can perform a second order polynomial expansion at a point $j$ about $i$: 

\begin{eqnarray}
q_{ij}  & = & a_{0,i} + a_{1,i} x_{ij} + a_{2,i} y_{ij} + a_{3,i} z_{ij} + a_{4,i} x_{ij}^2 + a_{5,i} y_{ij}^2 + \nonumber \\
& &  a_{6,i} z_{ij}^2 + a_{7,i} x_{ij} y_{ij} + a_{8,i} x_{ij}z_{ij} + a_{9,i} y_{ij} z_{ij}  + \nonumber \\
& & O(h^3)
\end{eqnarray}
where ${\bf x_{ij}} = \urij / h_i = [x_{ij}, y_{ij}, z_{ij}]$. 

The coefficients of this expansion can then be determined by inverting the following $10\times10$ matrix equation:

\begin{equation}
{\bf M} {\bf a} = {\bf q} 
\label{eqn:matrixeq}
\end{equation}  
where:
\begin{equation}
{\bf a}^\mathrm{T} = [a_0, a_1, a_2, a_3, a_4, a_5, a_6, a_7, a_8, a_9] 
\end{equation}
\begin{eqnarray}
{\bf q}^\mathrm{T} & = \sum_j^N m_j q_j \overline{W}_{ij} & \left[1, x_{ij}, y_{ij}, z_{ij},x_{ij}^2,y_{ij}^2,z_{ij}^2, \right . \nonumber \\
& & \left . x_{ij}y_{ij},x_{ij}z_{ij},y_{ij}z_{ij}\right]
\end{eqnarray}
\begin{eqnarray}
{\bf M} & = & \sum_j^N m_j \overline{W}_{ij}
\left(\begin{array}{rrr}
1 & \dx & \dy \cdots \\
\dx & \dx^2 & \dx\dy \cdots \\
\dy  & \dy \dx & \dy^2 \cdots \\
\dz & \dz \dx & \dz \dy \cdots \\
\dx^2 & \dx^3 & \dx^2\dy \cdots \\
\dy^2 & \dy^2 \dx & \dy^3 \cdots \\
\dz^2  & \dz^2 \dx & \dz^2 \dy \cdots \\
\dx\dy & \dx^2\dy & \dx\dy^2 \cdots \\
\dx\dz  & \dz\dx^2 & \dx\dz\dy \cdots \\
\dy\dz & \dy\dz \dx & \dz \dy^2 \cdots \\
 \end{array} \right . \nonumber \\ 
& &  \left . \begin{array}{lrrr}
 \cdots \dz & \dx^2 & \dy^2 & \dz^2 \cdots \\
\cdots  \dx\dz & \dx^3 & \dx\dy^2 & \dx\dz^2 \cdots\\
\cdots \dy\dz & \dy\dx^2 & \dy^3 & \dy\dz^2  \cdots\\
\cdots \dz^2 & \dz \dx^2 & \dz \dy^2 & \dz^3 \cdots\\
\cdots \dx^2\dz & \dx^4 & \dx^2\dy^2 & \dx^2\dz^2 \cdots \\
\cdots \dy^2 \dz & \dy^2 \dx^2 & \dy^4 & \dy^2 \dz^2 \cdots\\
\cdots \dz^3 & \dz^2 \dx^2 & \dz^2 \dy^2 & \dz^4  \cdots\\
\cdots \dx\dy\dz & \dy\dx^3 & \dx\dy^3 & \dx\dy\dz^2  \cdots\\
\cdots \dx\dz^2 & \dz\dx^3 & \dx\dz\dy^2 & \dx\dz^3 \cdots \\
\cdots \dy \dz^2 & \dy\dz \dx^2 & \dz \dy^3 & \dy \dz^3  \cdots\\
 \end{array} \right . \nonumber \\ 
& &  \left . \begin{array}{lrr}
\cdots \dx \dy & \dx \dz & \dy \dz \\
\cdots \dx^2 \dy & \dx^2 \dz & \dx \dy \dz \\
\cdots \dx \dy^2 & \dy\dx \dz & \dy^2 \dz \\
\cdots \dz \dx \dy & \dx \dz^2 & \dy \dz^2 \\
\cdots \dx^3 \dy & \dx^3 \dz & \dx^2\dy \dz \\
\cdots \dx \dy^3 & \dy^2 \dx \dz & \dy^3 \dz \\
\cdots \dz^2 \dx \dy & \dx \dz^3 & \dy \dz^3 \\
\cdots \dx^2 \dy^2 & \dy\dx^2 \dz & \dx\dy^2 \dz \\
\cdots \dz\dx^2 \dy & \dx^2 \dz^2 & \dx\dy \dz^2 \\
\cdots \dz \dx \dy^2 & \dy \dx \dz^2 & \dy^2 \dz^2 \\
\end{array} \right)
\label{eqn:M3d}
\end{eqnarray}
and $\overline{W}_{ij} = \frac{1}{2}[W_{ij}(h_i) + W_{ij}(h_j)]$ is the symmetrised smoothing kernel (the superscript $^\mathrm{T}$ means transpose). 

Having determined all of the coefficients of ${\bf a}$ (by solving ${\bf a} = {\bf M}^{-1} {\bf q}$), the gradients of $q$ evaluated at $i$ then simply follow as: 

\begin{equation}
\frac{\partial q_i}{\partial x} = a_1 ; \frac{\partial q_i}{\partial y} = a_2 ;\frac{\partial q_i}{\partial z} = a_3
\end{equation}
and similarly for the second derivatives. 

\section{A general derivation of conservation terms for multimass \SPHG}\label{sec:massdisscons} 

In section \ref{sec:diffm}, we introduced a multimass dissipation term for \SPHG. This requires some correction terms to restore energy and momentum conservation. In this appendix, we derive the general class of such correction terms. 

Our dissipation terms must obey mass, momentum and energy conservation: 

\begin{equation}
\dot{M} = 0 = \sum_j \dot{m}_j 
\label{eqn:mcons}
\end{equation}
\begin{equation}
\dot{M{\bf V}} = 0 = \sum_j \dot{m_j {\bf v}_j} = \sum_j m_j \dot{\bf v}_j + \dot{m}_j {\bf v}_j 
\label{eqn:fullmomcons}
\end{equation}
\begin{equation}
\dot{E} = 0 = \sum_j \dot{m}_j \left(\frac{1}{2}{\bf v}_j \cdot {\bf v}_j + u_j\right) + m_j\left({\bf v}_j\cdot \dot{{\bf v}}_j + \dot{u}_j\right)
\label{eqn:econsfull}
\end{equation}
First, let us verify that equation \ref{eqn:mdiss} satisfies equation \ref{eqn:mcons}. Substituting for $\dot{m}_j = \dot{m}_{\mathrm{diss},j}$, we have: 
\begin{equation}
\dot{M} = \sum_{j,k} Q_{jk} (m_j-m_k) K_{jk} = 0 
\end{equation}
where $Q_{ij} = Q_{ji} = \overline{m}_{ij}/\overline{\rho}_{ij} \overline{\alpha}_{ij} v^p_{\mathrm{sig},ij}L_{ij}$ and the above is zero because it is antisymmetric in $j,k$. Note that this explains why we must use a symmetrised mass in equation \ref{eqn:mdiss}: $Q_{ij}$ must be symmetric in order to ensure mass conservation.

Now, let us substitute $\dot{m}_j = \dot{m}_{\mathrm{diss},j}$ into equation \ref{eqn:fullmomcons}: 

\begin{equation}
0 = \sum_j m_j \dot{{\bf v}}_{\mathrm{diss},j} + \sum_{j,k} Q_{jk} (m_j - m_k) K_{jk} {\bf v}_j
\label{eqn:vdissconstr}
\end{equation}
where we have split the acceleration into a dissipative correction term, and all other normal \SPHG\ terms: $\dot{\bf v}_i = \dot{{\bf v}}_{\mathrm{diss},i} + \dot{{\bf v}}_{\mathrm{rest},i}$, and then used the fact that $\sum_j m_j \dot{{\bf v}}_{\mathrm{rest},j} = 0$ by construction for \SPHG. 

We may now select any form we like for $\dot{{\bf v}}_{\mathrm{diss},i}$ so long as it satisfies equation \ref{eqn:vdissconstr}. In \S\ref{sec:diffm}, we chose a form that conserves momentum on a per particle basis, but we may also choose a form that fluxes the momenta, for example: 

\begin{equation}
\dot{{\bf v}}_i = \sum_j Q_{ij} \frac{(m_i - m_j)}{m_i} K_{ij} {\bf v}_j
\label{eqn:methodbmom}
\end{equation}
It is straightforward to show that the above correction term also conserves momentum since it makes equation \ref{eqn:vdissconstr} antisymmetric in $j,k$. 

We may then derive a similar constraint equation for our energy correction term. As an example, let us substitute $\dot{m}_j = \dot{m}_{\mathrm{diss},j}$ and equation \ref{eqn:methodbmom} into equation \ref{eqn:econsfull}. This gives: 

\begin{eqnarray}
0 & = & \sum_{j,k} \left(Q_{jk} (m_j - m_k) K_{jk}\right)\left[\frac{1}{2}{\bf v}_j \cdot {\bf v}_j + u_j\right] +\nonumber \\ 
&  &  m_j \left[{\bf v}_j \cdot\left(Q_{jk} \frac{(m_j - m_k)}{m_j} K_{jk} {\bf v}_k\right) + \dot{u}_{\mathrm{diss},j}\right]
\label{eqn:engconsjk}
\end{eqnarray}
where similarly to the above, we have dropped all contributions involving the standard \SPHG\ terms since these are already conservative and therefore vanish. 

We may then derive a correction term for $\dot{u}_{\mathrm{diss},j}$: 

\begin{equation}
\dot{u}_{\mathrm{diss},i} = \sum_j Q_{ij} \frac{(m_i - m_j)}{m_i} K_{ij} \left[\frac{1}{2}{\bf v}_j \cdot {\bf v}_j + u_j\right] 
\label{eqn:methodbeng}
\end{equation}
It is straightforward to verify that substituting equation \ref{eqn:methodbeng} into equation \ref{eqn:engconsjk} makes the equation antisymmetric in $j,k$ and thus restores energy conservation. 

It is clear from the above examples that we may use the above constraints to derive a whole class of correction terms. Some of these may give better performance than equations \ref{eqn:masscons} and \ref{eqn:econs} that we use as our default in this paper. Such a study is, however, beyond the scope of this present work.  

\section{The importance of dissipation terms for multiphase and multimass flow in \SPHG}\label{sec:khdiss} 

\begin{center}
\begin{figure*}
\hspace{-6mm}\includegraphics[width=\textwidth]{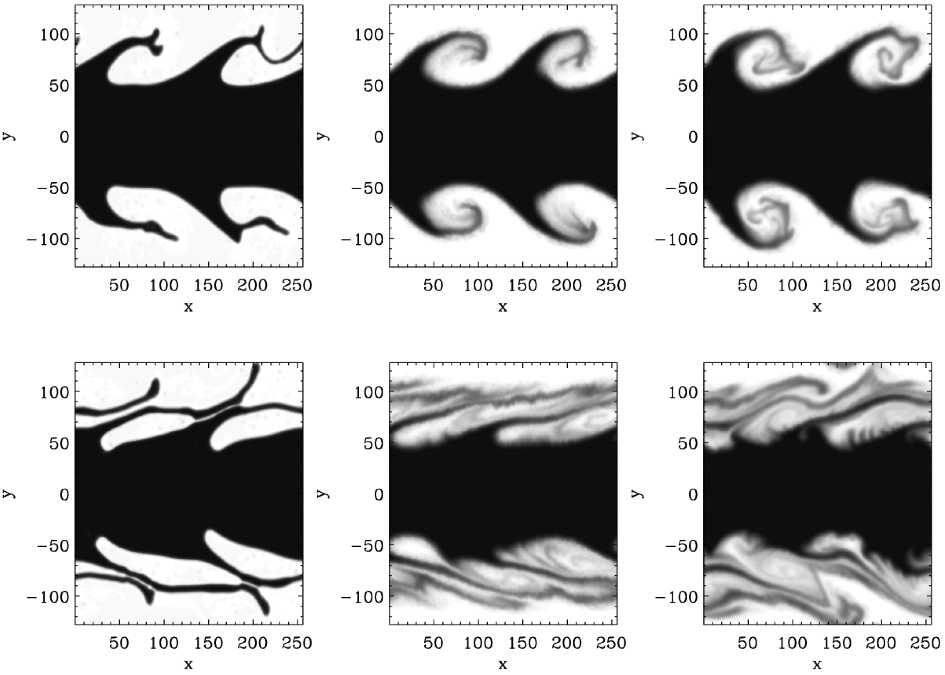}
\caption{KH1:8 multimass test results: the effect of removing the dissipation terms. From left to right the panels show: logarithmic density contours at: $\tau_\mathrm{KH}=1$ (top) and $\tau_\mathrm{KH}=2$ (bottom) for \SPHG\ run without entropy or mass dissipation (left); without mass dissipation (middle); and the full \SPHG\ scheme (right). (The right panels reproduce the results from Figure \ref{fig:kh}.)}
\label{fig:khdiss}
\end{figure*}
\end{center}

In this Appendix, we show the effect of switching off our dissipation terms in entropy and mass for the KH1:8 multimass test (\S \ref{sec:kh}). The results are shown in Figure \ref{fig:khdiss}. As expected, the entropy dissipation is extremely important: without it there is no mixing between the different fluid phases (see left panels). Notice, however, that even without dissipation, the KH rolls do grow on the correct timescale unlike in the classic SPH simulation (Figure \ref{fig:kh}, top row). This demonstrates (similarly to our findings in \Ron) that the improved force accuracy in \SPHG\ is responsible for the correct growth rate of the rolls, while the improved dissipation is responsible for actual mixing between the different fluid phases. 

The effect of the mass dissipation is more subtle. Without mass dissipation, mixing is also inhibited, but the effects are less strong than for the case without entropy dissipation because, unlike the entropies, the masses are smoothed inside the density sum\footnote{In fact, the results for this test without mass dissipation are rather similar to the KH1:8 single mass test we presented using OSPH in \Ron. This similarity arises because in OSPH the entropy -- like the particle masses -- is smoothed inside the pressure estimator (equation \ref{eqn:rtp}).} (equation \ref{eqn:sphcont}).

\section{The sensitivity of \SPHG\ to the dissipation parameters}\label{sec:alphasen} 

\begin{center}
\begin{figure*}
\hspace{-2mm}\includegraphics[width=\textwidth]{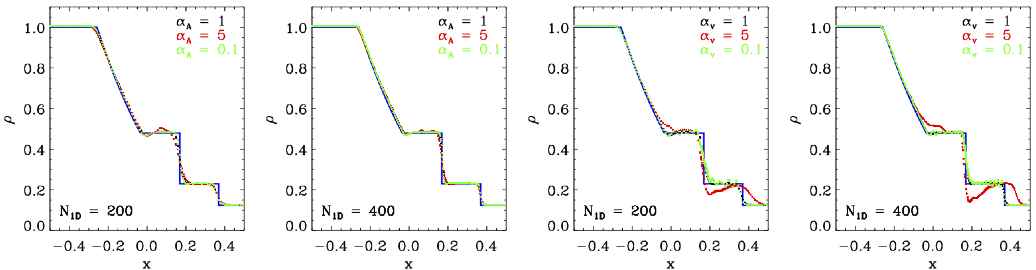}
\caption{Sensitivity to the dissipation parameters for the Sod shock tube test (\S\ref{sec:sod}). From left to right, the panels show the density profile at $t = 0.2$ (similarly to Figure \ref{fig:sod}), for varying entropy function dissipation parameter $\alpha_A$, and viscous dissipation parameter $\alpha_v$, at low resolution $N_{1D} = 200$ and higher resolution $N_{1D} = 400$, as marked. The blue line marks the analytic solution. Notice that so long as $\alpha_v$ is not too large, the results converge with increasing resolution independently of the choice of dissipation parameters.}
\label{fig:soddiss}
\end{figure*}
\end{center}

In this appendix, we assess how sensitive \SPHG\ is to the choice of dissipation parameters. As our default, we have assumed a single dissipation parameter for viscosity, mass dissipation and entropy dissipation: $\alpha = \alpha_v = \alpha_m = \alpha_A = 1$. This default choice is natural from the definition of the dissipation/viscosity equations \ref{eqn:viscmom}, \ref{eqn:visc}, \ref{eqn:entdiss} and \ref{eqn:mdiss}. These assert that the dissipation should proceed proportional to the jump in the given fluid quantity (mass, entropy etc.) and on a timescale set by the signal velocity. Thus, we expect a normalisation parameter in each case of order unity. Nonetheless, $\alpha$ is a free parameter and we should check that our results are not sensitive to it. To test this, in Figure \ref{fig:soddiss}, we consider the effect of varying $\alpha_v$ and $\alpha_A$ for the Sod shock tube test (\S\ref{sec:sod}) at two different resolutions. 

From Figure \ref{fig:soddiss}, we see that our results are not sensitive to the entropy dissipation parameter $\alpha_A$. Over a wide range $0.1 < \alpha_A < 5$, the results change only slightly. More importantly, the differences {\it decrease} with increasing resolution (compare the red and black lines in the left two panels of Figure \ref{fig:soddiss}). The results are more sensitive, however, to the choice of viscosity parameter $\alpha_v$. For low viscosity ($\alpha_v = 0.1$), we have spurious oscillations in the solution. Reassuringly, however, these decrease with increasing resolution (compare the green and black curves in the right two panels of Figure \ref{fig:soddiss}). The results are poor, however, if $\alpha_v$ is too large. For $\alpha_v = 5$, there is a strong under-shoot in the density at the shock that does not improve with increasing resolution. 

We conclude that the results in \SPHG\ converge with increasing resolution independently of $\alpha_v$ or $\alpha_A$, so long as $\alpha_v$ is not too large.

\bibliographystyle{mn2e}
\bibliography{../../BibTeX/refs}

\end{document}